\documentclass[aps,prb,twocolumn,superscriptaddress,longbibliography]{revtex4-2}

\usepackage{amsmath,amssymb,amsfonts}
\usepackage{graphicx}
\usepackage{dcolumn}
\usepackage{bm}
\usepackage{hyperref}
\usepackage{xcolor}
\usepackage{physics}
\usepackage{bbm}
\usepackage{mathtools}
\usepackage{booktabs}

\newcommand{\kB}{k_{\rm B}}

\newcommand{\mxy}{\mathcal{M}_{xy}}
\newcommand{\PT}{\mathcal{PT}}
\newcommand{\CfourT}{C_{4z}\mathcal{T}}

\newcommand{\sigxy}{\sigma_{xy}}
\newcommand{\axy}{\alpha_{xy}}
\newcommand{\kxy}{\kappa^{(0)}_{xy}}
\newcommand{\sigxys}{\sigma^{s}_{xy}}
\newcommand{\axys}{\alpha^{s}_{xy}}
\newcommand{\kxys}{\kappa^{(0),s}_{xy}}

\newcommand{\Lo}{L_0}
\newcommand{\jone}{j_1}
\newcommand{\jtwo}{j_2}

\begin{document}

\title{Light-Induced Topological Phase Transitions and Anomalous Thermal Transport in d-Wave Altermagnets }

\author{Ayesha Maryam}
\affiliation{Department of Physics, Quaid i Azam University, Islamabad, Pakistan}

\author{Muzamil Shah}
\email{muzamil@qau.edu.pk}
\affiliation{Department of Physics, Quaid i Azam University, Islamabad, Pakistan}
\affiliation{Research Center of Astrophysics and Cosmology, Khazar University, Baku AZ1096, 41 Mehseti Street, Azerbaijan}
\author{Kashif Sabeeh}
\email{kashifsabeeh@hotmail.com}
\affiliation{Department of Physics, Quaid i Azam University, Islamabad, Pakistan}

\author{Reza Asgari}
\affiliation{Department of Physics, Zhejiang Normal University, Jinhua 321004, P. R. China}
\affiliation{School of Quantum Physics and Matter, Institute for Research in Fundamental Sciences (IPM), Tehran 19395-5531, Iran}

\date{\today}

%=============================================================
\begin{abstract}
We study intrinsic thermal transport and Floquet-engineered topology in a two-dimensional d-wave altermagnetic topological insulator powered by linearly polarized light. We analyze the anomalous Hall, Nernst, and thermal Hall conductivities, as well as their spin-resolved equivalents, and develop closed-form formulas for the Berry curvature using an analytically calculated high-frequency effective Hamiltonian. We demonstrate that linearly polarized light, in contrast to conventional antiferromagnets, breaks the symmetry connecting spin sectors in altermagnets, allowing a series of spin-selective topological phase transitions from a quantum spin Hall state to a spin-polarized Chern insulator and finally to a trivial phase. The Nernst response shows substantial thermal activation and significant sensitivity to the gap size in the Chern domain, but both the electrical and thermal Hall responses become quantized and meet the anomalous Wiedemann–Franz law. Every anomalous transport coefficient exhibits a distinctive d-wave dependence on the polarization angle, reversing sign under orthogonal rotation and vanishing at symmetry-restoring directions. Our findings show a path to all-optical regulation of topological and caloritronic responses beyond traditional magnetic systems and establish thermal transport as a sensitive probe of altermagnetic order.

\end{abstract}

\maketitle

%=============================================================
\section{Introduction}
\label{sec:intro}
%=============================================================

Altermagnetism (AM) has recently emerged as a distinct magnetic phase characterized by zero net magnetization coexisting with strongly anisotropic, momentum-dependent spin splitting~\cite{PhysRevX.12.040501,PhysRevLett.117.090402,schellekens2014ultrafast, Mazin2022, song2025altermagnets}, and has been identified in a growing number of materials~\cite{Ahn2019, Yuan2020, Hayami2019, Mazin2021, Li2025}. Unlike conventional antiferromagnets (AFMs), which preserve combined parity–time ($\PT$) symmetry and exhibit spin-degenerate bands, altermagnets break both time-reversal ($\mathcal{T}$) and $\PT$ symmetries while preserving their combination with specific crystal rotations~\cite{liu2026linearly}. This distinctive symmetry structure~\cite{jungwirth2026symmetry} enables a wide range of spintronic and topological phenomena, including giant magnetoresistance~\cite{Smejkal2022b}, anomalous Hall effects~\cite{attias2024intrinsic}, topological insulating phases~\cite{ma2024altermagnetic, gonzalez2025spin, gonzalez2025model}, and unconventional superconducting proximity effects~\cite{zhang2024finite}.

Light is a particularly powerful external stimulus for controlling quantum materials, as it couples directly to electronic degrees of freedom and can dynamically modify band structure and symmetry. Beyond enabling control of intrinsic electronic and magnetic properties, time-periodic driving can induce nonequilibrium phases that have no static counterparts. Within the framework of Floquet theory~\cite{lindner2011floquet, oka2019floquet, rudner2020band}, such periodic driving provides a versatile route to engineer effective Hamiltonians and realize dynamically generated topological states~\cite{eckardt2015high,Mikami2016BWFloquet}.

In conventional nonmagnetic and antiferromagnetic systems, Floquet engineering has enabled a wide range of phenomena, including light-induced Hall effects~\cite{oka2009photovoltaic,mciver2020light}, Floquet topological insulators~\cite{ma2024altermagnetic, lindner2011floquet, else2016floquet}, light-induced Weyl semimetals~\cite{PhysRevLett.132.016603}, and Floquet time crystals~\cite{PhysRevLett.117.090402}. It has also opened a route to ultrafast control of spin and magnetic textures on picosecond and sub-picosecond timescales, forming the basis of ultrafast spintronics~\cite{schellekens2014ultrafast,rudner2020band}. In topological systems, light-driven Floquet states have enabled novel phenomena such as Majorana modes, light-controlled edge transport, and dynamical probes of topological phase transitions.

Recently, the effects of light on unconventional magnetic systems, particularly altermagnets, have begun to attract significant attention~\cite{liu2026linearly,Zhu2025FloquetOddParity,Ghorashi2025DynamicalGeneration}. However, existing studies have largely focused on charge transport and high-frequency regimes.
While Berry-curvature-driven thermal transport 
is well established in topological 
materials~\cite{nagaosa2010anomalous}, its Floquet 
engineering in altermagnetic systems remains 
essentially unexplored. This is especially 
important because symmetry considerations 
allow linearly polarized light (LPL) to 
induce Hall responses in altermagnets, in 
contrast to conventional antiferromagnets.

More broadly, recent work has demonstrated 
that Berry-curvature-related responses in 
low-dimensional and compensated magnetic 
systems can be tuned by a variety of external 
control parameters beyond optical driving, 
including strain, electric fields, and 
heterostructure engineering. Notable recent 
advances include the demonstration of 
externally controlled anomalous Hall and 
spin-dependent transport responses in 
low-dimensional materials~\cite{chen2025strain} 
and compensated magnetic 
systems~\cite{shen2026dual}. Our work 
contributes to this broader direction by 
establishing linearly polarized light as an 
all-optical control knob for anomalous 
thermal and spin-caloritronic responses in 
altermagnetic topological insulators, without 
requiring magnetic field reversal or 
structural modification.

In nonmagnetic and conventional antiferromagnetic systems, LPL generally cannot generate a Hall response because $\PT$ symmetry is preserved. Altermagnets circumvent this restriction: their intrinsically broken $\PT$ symmetry allows LPL, by breaking the spin-rotation symmetry $\CfourT$ that relates the two spin sublattices, to induce a finite anomalous Hall effect~\cite{liu2026linearly}. This mechanism is part of a wider effort to 
identify external control parameters that can 
tune Berry-curvature-driven responses in 
systems with compensated magnetic order, 
where conventional Hall probes are 
ineffective~\cite{chen2025strain, shen2026dual}. 
Thermal and spin-caloritronic channels, as 
studied here, offer particularly sensitive 
probes of such responses precisely because 
they weight the Berry curvature differently 
from charge transport. Recent works have demonstrated LPL-induced anomalous Hall responses and topological phase transitions in altermagnetic topological insulators~\cite{liu2026linearly}, as well as circularly polarized light–induced quantum anomalous Hall phases~\cite{Zou2025FloquetQAH}, light-induced odd-parity magnetism~\cite{Zhu2025FloquetOddParity,Zou2025FloquetQAH}, and higher-order spin–orbit coupling effects~\cite{Ghorashi2025DynamicalGeneration}.

Despite this progress, a comprehensive understanding of thermal transport under LPL irradiation in altermagnetics remains lacking. Thermal transport coefficients; namely the anomalous Nernst conductivity (ANC) and anomalous thermal Hall conductivity (ATHC), probe the Berry curvature through distinct thermal weightings that differ fundamentally from the anomalous Hall conductivity (AHC). As a result, they provide complementary and often more sensitive insights into the underlying topology and magnetic order. The ANC, being thermally activated, serves as a sensitive probe of band topology at finite temperature, while the ATHC, together with its relation to $\sigxy$ via the Wiedemann–Franz (WF) law, offers a direct test of topological quantization. Furthermore, the spin-resolved counterparts; the spin Nernst and spin thermal Hall conductivities, enable access to spin caloritronic effects of direct relevance for device applications~\cite{Kitagawa2011}.

In this work, we present a comprehensive theoretical study of thermal transport in a $d$-wave altermagnetic  under LPL irradiation. Starting from an analytically derived Floquet effective Hamiltonian in the high-frequency limit, we obtain explicit expressions for the Berry curvature in each spin sector and compute six intrinsic transport coefficients: the AHC $\sigxy$, ANC $\axy$, ATHC $\kxy$, as well as their spin-resolved counterparts $\sigxys$, $\axys$, and $\kxys$. We analyze their dependence on topological phase, polarization angle, and temperature, and establish their interrelations via the Mott and Wiedemann–Franz laws. Finally, we contrast the altermagnetic response with that of conventional antiferromagnets, identifying clear experimental signatures accessible through thermal Hall and Nernst measurements.

%=============================================================
\section{Model and Floquet Effective Hamiltonian}
\label{sec:model}
We employ the four-band two-dimensional out-of-plane $d$-wave
altermagnetic model on a square lattice~\cite{gonzalez2025spin,PhysRevB.110.064426,liu2026linearly} in which the orbitals are $p_{z\uparrow}$, $\frac{1}{\sqrt{2}}(d_{xz\uparrow} + i d_{yz\uparrow})$, $p_{z\downarrow}$, and $\frac{1}{\sqrt{2}}(d_{xz\downarrow} - i d_{yz\downarrow})$.  
The low-energy Hamiltonian has the block-diagonal form
\begin{equation}
H(\bm{k}) =
\begin{pmatrix} H_\uparrow(\bm{k}) & 0 \\ 0 & H_\downarrow(\bm{k}) \end{pmatrix},
\label{eq:H0}
\end{equation}
where each $2\times 2$ spin block is written in terms of a $d$-vector as
$H_\sigma(\bm{k}) = \bm{d}^\sigma(\bm{k})\cdot\bm{\tau}$, with
$\bm{\tau}=(\tau_x,\tau_y,\tau_z)$ the Pauli matrices acting on the
orbital subspace.
The explicit $d$-vector components are
\begin{align}
d^\uparrow_x &= -v\sin k_x,\quad
d^\uparrow_y = -vt_a\sin k_y, \nonumber\\
d^\uparrow_z &= m + b(\cos k_x + t_a^2\cos k_y),
\label{eq:dup}
\end{align}
and $H_\downarrow(k_x,k_y)=H_\uparrow(k_y,k_x)$, i.e., the spin-down
block is obtained by the exchange $k_x\leftrightarrow k_y$.
Here $m$ is the on-site potential, $v$ the hopping amplitude,
$b$ the inversion-breaking hopping, and $t_a$ the asymmetry parameter.
The two spin channels are related by $\CfourT$ or mirror symmetry
$\mxy$. The lattice constant is assumed to be unity. 
Setting $t_a=1$ restores $\PT$ symmetry, reducing the system to a
conventional AFM; $t_a\neq 1$ breaks $\PT$ and yields the $d$-wave AM.
The equilibrium system is a QSH insulator when
\begin{equation}
|m| < |b|(1 + t_a^2),
\label{eq:qsh-condition}
\end{equation}
and topologically trivial otherwise~\cite{ma2024altermagnetic}. 
The parameters used throughout this work,
\begin{equation}
m = 3.8v, \qquad b = -v, \qquad t_a = \sqrt{3},
\label{eq:parameters}
\end{equation}
satisfy this condition since
\begin{equation}
|b|(1 + t_a^2) = v(1 + 3) = 4v > 3.8v = m,
\label{eq:qsh-check}
\end{equation}
placing the system in the QSH regime but 
sufficiently close to the topological phase 
boundary at $m = 4v$ that light-induced 
renormalization of the hopping parameters can 
drive sequential gap closings in the two spin 
sectors. The hopping parameter $v$ sets the 
overall energy scale of the model, the ratio 
$m/b$ controls the degree of band inversion, 
and $t_a \neq 1$ breaks the combined 
parity--time ($\mathcal{PT}$) symmetry that 
would otherwise reduce the system to a 
conventional antiferromagnet.

We drive the system with LPL whose vector potential is given as
\begin{equation}
\bm{A}(t) = A_0[\cos\theta\cos(\omega t),\,\sin\theta\cos(\omega t)],
\end{equation}
where $A_0$ is the amplitude (it is dimensionless in our units), $\theta$ the polarization angle relative
to the $x$-axis, and $\omega$ the frequency.
The time-dependent Hamiltonian is obtained via the Peierls substitution
$\bm{k}\to\bm{k}+e\bm{A}(t)/\hbar$.
In the high frequency limit and off-resonant regime, the leading order commutator correction
vanishes because $\bm{A}(t)=\bm{A}(-t+\tau)$ for a fractional time
translation $\tau$, and the effective Hamiltonian reduces to the time averaged part~\cite{eckardt2015high} in Floquet high-frequency expansion regime.

To derive effective hamiltonan explicitly we 
substitute $\mathbf{k} \to \mathbf{k} + 
e\mathbf{A}(t)/\hbar$ into the static $d$-vector 
of Equation~\eqref{eq:dup} For the spin-up block this gives, 
for example,
\begin{equation}
d^{\uparrow}_x(t) = -v\sin\!\left(k_x + 
\frac{eA_0\cos\theta}{\hbar}\cos\omega t\right),
\end{equation}
and similarly for $d^{\uparrow}_y(t)$ and $d^{\uparrow}_z(t)$, with the driving entering along $k_x$ through $A_0\cos\theta$ and along $k_y$ through $A_0\sin\theta$. In the high-frequency off-resonant regime, the leading correction to the effective Hamiltonian is given by the time average over one driving period,
\begin{equation}
H^{\sigma}_{\mathrm{eff}}(\mathbf{k}) = 
\frac{1}{T_0}\int_0^{T_0} H^{\sigma}(\mathbf{k},t)\,dt,
\qquad T_0 = \frac{2\pi}{\omega}.
\end{equation}
Expanding $\sin(k_x + \phi(t))$ and 
$\cos(k_x + \phi(t))$ using standard 
trigonometric identities and performing the time 
average using the Jacobi--Anger expansion,
\begin{equation}
\frac{1}{T_0}\int_0^{T_0} 
\cos\!\left(\frac{eA_0\cos\theta}{\hbar}
\cos\omega t\right) dt 
= J_0(A_0\cos\theta) \equiv j_1,
\end{equation}
(and analogously for the $k_y$-dependent terms, 
giving $j_2 = J_0(A_0\sin\theta)$), one obtains 
directly the renormalized $d$-vector components 
in Eqs.~\eqref{eq:deff_up} and ~\eqref{eq:deff_dn}: every term that originally 
depended on $k_x$ acquires a factor $j_1$, and 
every term depending on $k_y$ acquires a factor 
$j_2$. The terms odd in the driving field 
(generating Bessel functions $J_{n\geq1}$) 
average to zero in this leading-order expansion, 
leaving only the $J_0$ renormalization factors 
$j_1$ and $j_2$.

Physically, $j_1$ and $j_2$ represent the 
LPL-induced renormalization of the effective 
hopping amplitudes along the $x$- and 
$y$-directions, respectively. Because the 
spin-down block is related to the spin-up block 
by the coordinate exchange $k_x \leftrightarrow 
k_y$ [Equation~\eqref{eq:dup}], the same substitution shows that 
$H_{\downarrow,\mathrm{eff}}$ is obtained from 
$H_{\uparrow,\mathrm{eff}}$ by simultaneously 
exchanging $k_x \leftrightarrow k_y$ and 
$j_1 \leftrightarrow j_2$, which is precisely 
the origin of Equation~\eqref{eq:deff_up}. When $\theta = \pm\pi/4$, 
$j_1 = j_2$ and the two spin blocks remain related 
by the original symmetry operation; for any other 
$\theta$, $j_1 \neq j_2$ and this symmetry is 
broken, which is the microscopic origin of the 
spin-selective topological phase transitions 
discussed throughout this work.

In the off-resonant limit, the heating rate is significantly reduced, allowing the system to reach a long-lived Floquet prethermal \cite{machado2019exponentially, ho2023quantum} state where the effective Hamiltonian description remains valid or the situation where a finite pulse duration in order of ps or ns is used. 

Finite temperature primarily broadens the Fermi 
distribution but does not destroy the topological 
band structure, provided $k_BT \ll \hbar\omega$. 
The high-frequency expansion underlying 
Eqs.~\eqref{eq:deff_up} and ~\eqref{eq:deff_dn} is controlled by the parameter 
$W/\hbar\omega$, where $W \sim |v| + |b|(1+t_a^2)$ 
is the electronic bandwidth of the static model. 
Truncation at leading order in this expansion, 
as performed here, is justified when
\begin{equation}
\hbar\omega \gg W.
\label{eq:high-freq-condition}
\end{equation}
For typical hopping parameters $v, |b| \sim 
0.1$--1~eV, the bandwidth is $W \sim 0.5$--2~eV in 2D semiconductor materials. 
Driving frequencies in the mid-infrared to 
terahertz range, $\omega/2\pi \sim 10$--100~THz 
(i.e., $\hbar\omega \sim 0.04$--0.4~eV), 
therefore satisfy Eq.~(\ref{eq:high-freq-condition}) 
only marginally for the larger end of $W$, but 
comfortably for systems with smaller bandwidths 
or for higher driving frequencies. We assume 
throughout that the system is operated in this 
regime, so that higher-order Floquet corrections 
do not qualitatively modify the topological phases 
discussed below.

Applying the Jacobi Anger expansion gives renormalized $d$-vectors:
\begin{align}
d^{\uparrow}_{x,\rm eff} &= -v\jone\sin k_x, \quad
d^{\uparrow}_{y,\rm eff} = -vt_a\jtwo\sin k_y, \nonumber\\
d^{\uparrow}_{z,\rm eff} &= m + b[\jone\cos k_x + \jtwo t_a^2\cos k_y],
\label{eq:deff_up}
\end{align}
\begin{align}
d^{\downarrow}_{x,\rm eff} &= -v\jtwo\sin k_y, \quad
d^{\downarrow}_{y,\rm eff} = -vt_a\jone\sin k_x, \nonumber\\
d^{\downarrow}_{z,\rm eff} &= m + b[\jtwo\cos k_y + \jone t_a^2\cos k_x],
\label{eq:deff_dn}
\end{align}
where
\begin{equation}
\jone = J_0(A_0\cos\theta), \quad \jtwo = J_0(A_0\sin\theta),
\label{eq:j12}
\end{equation}
and $J_0$ is the zeroth-order Bessel function.
The key physical content of Eqs.~\eqref{eq:deff_up}--\eqref{eq:j12}
is transparent: when $\theta\neq\pm\pi/4$, we have $\jone\neq\jtwo$,
which means the spin-up and spin-down effective Hamiltonians are no
longer related by any symmetry operation, breaking $\CfourT$ and
$\mxy$.
In contrast, for the conventional AFM ($t_a=1$), the $\PT$ symmetry is
preserved for any $\theta$, locking $\jone=\jtwo$ in its effect on the
combined system and hinders any Hall response.

The eigenvalues of $H^\sigma_{\rm eff}$ are
$\varepsilon^\sigma_\pm(\bm{k}) = \pm|\bm{d}^\sigma_{\rm eff}(\bm{k})|$.
A topological phase transition occurs whenever a spin-resolved gap
$\Delta_\sigma = \min_{\bm{k}}[2|\bm{d}^\sigma_{\rm eff}(\bm{k})|]$
closes. Since $\jone\neq\jtwo$ under LPL in the AM, the two gaps
$\Delta_\uparrow$ and $\Delta_\downarrow$ evolve independently with
$A_0$, enabling sequential closings that are forbidden in the AFM.

We emphasize that the parameter set in 
Eq.~(\ref{eq:parameters}) defines a 
symmetry-allowed minimal-model regime and 
should not be interpreted as a quantitative 
fit to any specific material. To our 
knowledge, an intrinsic QSH phase in a 
monolayer altermagnetic compound has not yet 
been unambiguously confirmed experimentally. 
Our model instead provides a minimal 
theoretical platform that isolates the role 
of altermagnetic symmetry and linearly 
polarized light in producing spin-selective 
topological phase transitions.

Several candidate material directions may 
nevertheless allow the required conditions 
to be realized:

MnTe-based heterostructures.
MnTe is a well-established altermagnetic 
candidate~\cite{vsmejkal2022beyond} with strong 
spin-momentum locking. Thin-film or 
heterostructure engineering could 
introduce the band inversion needed for 
a QSH phase while preserving the 
altermagnetic symmetry.
    
RuO$_2$-based thin films.
RuO$_2$ has been identified as a $d$-wave 
altermagnet~\cite{ahn2019antiferromagnetism} and its 
electronic structure can be tuned 
significantly by epitaxial strain and 
substrate choice, potentially bringing 
the system into a band-inverted regime.
    
Artificial lattice platforms.
Moir\'{e} superlattices and cold-atom 
optical lattices with engineered 
square-lattice geometry offer full 
tunability of the hopping parameters 
$v$, $b$, and $t_a$, making it possible 
to place the system directly into the 
QSH altermagnetic regime defined by 
Eq.~(\ref{eq:qsh-condition}).

Quantitative predictions for these specific 
systems would require material-specific 
first-principles input and are left for 
future work.

%=============================================================
\section{Berry Curvature}
\label{sec:berry}
%=============================================================
Throughout this work we adopt the following conventions 
for the spin-resolved and total Chern numbers. The 
spin-sector Chern numbers are defined as
\begin{equation}
C_{\uparrow} = \frac{1}{2\pi}
\int_{\mathrm{BZ}} d^2k\, \Omega^{z,\uparrow}_{-}(\mathbf{k}),
\qquad
C_{\downarrow} = \frac{1}{2\pi}
\int_{\mathrm{BZ}} d^2k\, \Omega^{z,\downarrow}_{-}(\mathbf{k}),
\end{equation}
the total Chern number is
\begin{equation}
C = C_{\uparrow} + C_{\downarrow},
\end{equation}
and the spin Chern number is
\begin{equation}
C_s = \frac{C_{\uparrow} - C_{\downarrow}}{2}.
\end{equation}
The three topological phases realized under LPL irradiation 
are then characterized as follows:
\begin{align}
\text{QSH phase:} \quad 
& C_{\uparrow} = -1,\quad C_{\downarrow} = +1,
\quad C = 0,\quad C_s = -1, \\
\text{Chern phase:} \quad 
& C_{\uparrow} = -1,\quad C_{\downarrow} = 0,
\quad C = -1,\quad C_s = -\tfrac{1}{2}, \\
\text{Trivial phase:} \quad 
& C_{\uparrow} = 0,\quad C_{\downarrow} = 0,
\quad C = 0,\quad C_s = 0.
\end{align}
The spin-up and spin-
down components in the Eq.~(\ref{eq:H0}) are decoupled, their Berry connections and Chern numbers
can be evaluated independently. For the lower band $\varepsilon^\sigma_- = -|\bm{d}^\sigma|$ of each
spin block, the normalized eigenstate in terms of the spherical
coordinates $\hat{d}^\sigma = (\sin\Theta\cos\Phi,
\sin\Theta\sin\Phi, \cos\Theta)$ of the effective $d$-vector is
\begin{equation}
|u^\sigma_-(\bm{k})\rangle =
\begin{pmatrix} -\sin(\Theta/2)\,e^{-i\Phi} \\ \cos(\Theta/2) \end{pmatrix}.
\label{eq:eigenstate}
\end{equation}
Computing the Berry connection
$\mathcal{A}^\sigma_\mu = i\langle u^\sigma_-|\partial_{k_\mu}|u^\sigma_-\rangle$
from Eq.~\eqref{eq:eigenstate} gives
\begin{equation}
\mathcal{A}^\sigma_\mu(\bm{k}) = \sin^2\!\left(\frac{\Theta}{2}\right)
\partial_{k_\mu}\Phi.
\label{eq:berry_conn}
\end{equation}
Taking the antisymmetric curl of Eq.~\eqref{eq:berry_conn} and
converting to the solid angle form, the $z$-component of the Berry
curvature for the lower band of spin $\sigma$ can be expressed as
\begin{equation}
\Omega^{z,\sigma}_-(\bm{k}) = \frac{1}{2}\hat{d}^\sigma\cdot
\left(\partial_{k_x}\hat{d}^\sigma\times\partial_{k_y}\hat{d}^\sigma\right)
= \frac{\bm{d}^\sigma\cdot(\partial_{k_x}\bm{d}^\sigma
\times\partial_{k_y}\bm{d}^\sigma)}{2|\bm{d}^\sigma|^3}.
\label{eq:berry_curv}
\end{equation}
Evaluating the triple product explicitly with the LPL renormalized
$d$-vectors~\eqref{eq:deff_up} and \eqref{eq:deff_dn}, we obtain the
closed form expression for the spin-up Berry curvature as:
\begin{widetext}
\begin{equation}
\begin{split}
\Omega^{z,\uparrow}_-(\bm{k}) =
\frac{
bv^2 t_a j_1^2 j_2 \sin^2 k_x\cos k_y
+ bv^2 t_a^3 j_1 j_2^2 \sin^2 k_y\cos k_x
}{2|\bm{d}^\uparrow_{\rm eff}|^3} 
+ \frac{
d^\uparrow_{z,\rm eff}\, v^2 t_a j_1 j_2\cos k_x\cos k_y
}{2|\bm{d}^\uparrow_{\rm eff}|^3},
\end{split}
\label{eq:Omup}
\end{equation}
and for spin-down:
\begin{equation}
\begin{split}
\Omega^{z,\downarrow}_-(\bm{k}) =
\frac{
-bv^2 t_a j_1 j_2^2 \sin^2 k_y\cos k_x
- bv^2 t_a^3 j_1^2 j_2 \sin^2 k_x\cos k_y
}{2|\bm{d}^\downarrow_{\rm eff}|^3} 
- \frac{
d^\downarrow_{z,\rm eff}\, v^2 t_a j_1 j_2\cos k_x\cos k_y
}{2|\bm{d}^\downarrow_{\rm eff}|^3}.
\end{split}
\label{eq:Omdn}
\end{equation}
\end{widetext}
Equations~\eqref{eq:Omup} and~\eqref{eq:Omdn} encode several
important symmetry properties.
First, when $j_1=j_2$ (i.e., $A_0=0$ or $\theta=\pm\pi/4$), we can
verify directly that $\Omega^{z,\uparrow}_-(\bm{k})=
-\Omega^{z,\downarrow}_-(\bm{k})$, so the total Berry curvature
$\Omega^{z,\uparrow}_-+\Omega^{z,\downarrow}_-=0$ at every $\bm{k}$,
consistent with the preserved $\CfourT$ symmetry.
Second, under $\PT$ symmetry (AFM, $t_a=1$, any $j_1,j_2$):
$\PT$ maps $\Omega^z_n(\bm{k})\to -\Omega^z_n(\bm{k})$, forcing
$\Omega^z_n(\bm{k})=0$ for all $\bm{k}$, which prevents any Hall
response.
Third, the symmetry $\sigma_{xy}(\pi/2-\theta)=-\sigma_{xy}(\theta)$
is encoded in Eqs.~\eqref{eq:Omup}--\eqref{eq:Omdn} through the
exchange $j_1\leftrightarrow j_2$ under $\theta\to\pi/2-\theta$.

%=============================================================
\section{Thermal Transport}
\label{sec:transport}
%=============================================================
Having calculated the Berry phase, we can calculate some thermal transport quantities like anomalous Hall conductivity, Nernst conductivity and thermal Hall conductivity. 
\subsection{Anomalous Hall conductivity}
The low-energy and effective Hamiltonian is block diagonal in spin space thus the spin $S_z$ is a good quantum number, the valence
bands split into spin-up and spin-down components. Therefore, we can define the spin current in the system. Within the semiclassical Boltzmann framework, the anomalous (intrinsic)
contribution to the transverse conductivity is~\cite{crepieux2001theory}
\begin{equation}
\sigxy = \frac{e^2}{\hbar}\sum_{\sigma=\uparrow,\downarrow}
\sum_{n=\pm}
\int_{\mathrm{BZ}}\frac{d^2 k}{(2\pi)^2}
\Omega^{z,\sigma}_{n}(\mathbf{k})\,f^{\sigma}_{n}
\label{eq:AHC}
\end{equation}

where $f(\varepsilon)=(e^{\varepsilon/\kB T}+1)^{-1}$ is the
Fermi--Dirac distribution and $\mu$ is the chemical potential.
At $T=0$ in a gapped insulating phase, this reduces to
$\sigxy = (e^2/h)\,C$ 

For the two-band model considered here, the Berry curvatures 
of the upper and lower bands satisfy 
$\Omega^{z,\sigma}_{+}(\mathbf{k}) = -\Omega^{z,\sigma}_{-}(\mathbf{k})$, 
so when the chemical potential lies in the gap at $T = 0$, 
only the lower band contributes and one recovers the 
familiar Chern-number formula. For a general chemical 
potential or at finite temperature, however, both bands 
must be retained to correctly capture the decay of 
$\sigma_{xy}$ as $E_F$ moves into the conduction band.

where $C=C_\uparrow+C_\downarrow$ is the total
Chern number~\cite{sheng2006quantum}.

\subsection{Anomalous Nernst conductivity}
The intrinsic anomalous Nernst conductivity (ANC), 
relating the transverse electric current to a 
longitudinal temperature gradient via 
$J_y = \alpha_{xy}(-\partial_x T)$, is given by 
the Berry curvature weighted by the fermionic 
entropy kernel~\cite{xiao2006berry,nagaosa2010anomalous}:
\begin{equation}
\alpha_{xy}
=
\frac{ek_B}{\hbar}
\sum_{\sigma=\uparrow,\downarrow}
\sum_{n=\pm}
\int_{\mathrm{BZ}}\frac{d^2 k}{(2\pi)^2}
\,\Omega^{z,\sigma}_{n}(\mathbf{k})\,
s\!\left(f^{\sigma}_{n}\right),
\label{eq:ANC}
\end{equation}
where
\begin{equation}
s(f) = -f\ln f - (1-f)\ln(1-f)
\label{eq:entropy-kernel}
\end{equation}
is the entropy carried by a single fermionic state, 
and $f^{\sigma}_{n} = \left(\exp[(\varepsilon^{\sigma}_{n}
(\mathbf{k})-\mu)/k_BT]+1\right)^{-1}$ is the 
Fermi--Dirac distribution.

The vanishing of $\alpha_{xy}$ at zero temperature 
follows immediately from this form. As $T \to 0$, 
the Fermi function approaches a step function, so 
every state is either fully occupied ($f = 1$) or 
fully empty ($f = 0$). Since
\begin{equation}
s(1) = 0, \qquad s(0) = 0,
\label{eq:entropy-limits}
\end{equation}
the integrand vanishes identically at every 
$\mathbf{k}$ point when $\mu$ lies inside the bulk 
gap, and therefore
\begin{equation}
\alpha_{xy}(T = 0) = 0.
\label{eq:anc-zero-T}
\end{equation}
This result is a direct consequence of the third 
law of thermodynamics: a completely filled or 
completely empty state carries no entropy, and the 
Nernst response is therefore purely thermally 
activated. No cancellation between upper and lower 
bands is required; the vanishing follows directly 
from $s(0) = s(1) = 0$.

We note that the entropy-kernel form in 
Eq.~(\ref{eq:ANC}) is equivalent to the logarithmic 
representation obtained via integration by parts,
\begin{equation}
s(f^{\sigma}_{n}) \;\longleftrightarrow\; 
\frac{1}{k_BT}
\int_{\varepsilon^{\sigma}_{n}}^{\infty}
f(\varepsilon' - \mu)\,d\varepsilon'
= \ln\!\left(1 + e^{(\mu - 
\varepsilon^{\sigma}_{n})/k_BT}\right),
\label{eq:ibp-relation}
\end{equation}
up to an overall factor of $k_BT$. The logarithmic 
form is convenient for numerical evaluation, but 
the apparent divergence of 
$\ln(1+e^{(\mu-\varepsilon)/k_BT}) \approx 
(\mu-\varepsilon)/k_BT$ for occupied bands at low 
temperature is cancelled exactly by the vanishing 
Fermi occupation factor in the full thermoelectric 
kernel. The entropy-kernel form makes this 
cancellation manifest and is therefore the 
physically transparent starting point.
In practice, the entropy kernel $s(f)$ and the 
logarithmic form $\ln(1+e^{(\mu-\varepsilon)/k_BT})$ 
are numerically equivalent up to the factor $k_BT$, 
and both are well-behaved for $k_BT > 0$. Our 
numerical calculations use the entropy-kernel form 
of Eq.~(\ref{eq:ANC}) directly, which avoids any 
ambiguity about the low-temperature limit and is 
manifestly consistent with $\alpha_{xy}(T=0) = 0$.

\subsection{Anomalous thermal Hall conductivity}
The ATHC, defined by $J^Q_y = -\kxy\,\partial_x T$, is given by
the $c_2$ function formula derived via Luttinger's gravitational
potential approach~\cite{luttinger1964theory,qin2011energy}:
\begin{equation}
\kxy = -\frac{\kB^2 T}{\hbar}\sum_{\sigma=\uparrow,\downarrow}
\sum_{n=\pm}
\int_{\mathrm{BZ}}\frac{d^2 k}{(2\pi)^2}
\Omega^{z,\sigma}_{n}(\mathbf{k})\,c_2\!\left(f^{\sigma}_{n}\right),
\label{eq:ATHC}
\end{equation}

where the function $c_2$ is defined as
\begin{equation}
c_2(f) = \int_0^f\!\left(\ln\frac{1-t}{t}\right)^2\!dt,
\quad c_2(0)=0,\quad c_2(1)=\frac{\pi^2}{3}.
\label{eq:c2}
\end{equation}
The value $c_2(1)=\pi^2/3$ can be established precisely using the
Dirichlet eta function: $c_2(1)=2\int_0^\infty u^2 e^u/(e^u+1)^2\,du
= 4\eta(2) = \pi^2/3$.
In the low-temperature limit with the chemical potential 
sitting inside the gap, the upper band is completely 
empty so $f^{\sigma}_{+} \to 0$ and $c_2(0) = 0$, 
while the lower band is completely filled so 
$f^{\sigma}_{-} \to 1$ and $c_2(1) = \pi^2/3$. 
The upper band therefore contributes nothing in this 
limit and Eq.~(\ref{eq:ATHC}) reduces to
\begin{equation}
\left.\frac{\kxy}{T}\right|_{T\to 0} =
-\frac{\pi^2\kB^2}{3\hbar}\cdot\frac{C}{2\pi}
= -\frac{\pi^2\kB^2}{3h}\,C.
\label{eq:ATHC_quant}
\end{equation}

\subsection{Wiedemann--Franz law for anomalous transport}
Comparing Eq.~\eqref{eq:ATHC_quant} with $\sigxy=e^2 C/h$ at $T=0$,
we obtain the anomalous Wiedemann-Franz law:~\cite{jonson1980mott}
\begin{equation}
\frac{\kxy}{\sigxy} = -\frac{\pi^2\kB^2}{3e^2}\,T \equiv -\Lo T,
\label{eq:WF}
\end{equation}
where $\Lo=\pi^2\kB^2/3e^2\approx 2.44\times10^{-8}$\,W\,$\Omega$\,K$^{-2}$
is the Lorenz number.
This relation holds exactly in any topological phase at $T\to 0$
and serves as an unambiguous experimental test of the Berry phase
origin of the thermal Hall response.

\subsection{Mott formula for the Nernst conductivity}
At low temperature $\kB T\ll\mu$, the Sommerfeld expansion gives~\cite{jonson1980mott}
\begin{equation}
\axy \approx -\frac{\pi^2\kB^2 T}{3e}\,\frac{\partial\sigxy(\varepsilon)}
{\partial\varepsilon}\bigg|_{\varepsilon=\mu},
\label{eq:Mott}
\end{equation}
the anomalous analogue of the Mott formula.
This low-temperature expression is fully consistent 
with Eq.~(\ref{eq:ANC}): expanding $s(f)$ to leading 
order in $k_BT/(\varepsilon - \mu)$ via the Sommerfeld 
expansion and integrating by parts recovers 
Eq.~(\ref{eq:Mott}) directly~\cite{jonson1980mott}. 
The Mott relation therefore provides an independent 
check that the entropy-kernel form correctly reduces 
to zero at $T = 0$ and grows linearly in $T$ at low 
temperatures, consistent with the thermally activated 
picture established in Eq.~(\ref{eq:anc-zero-T}).
Equation~\eqref{eq:Mott} shows that $\axy$ is proportional to the
energy derivative of the AHC, making it sensitive to the sharpness
of Berry curvature features near the Fermi level.

\subsection{Spin-resolved transport coefficients}
Since the altermagnetic effective Hamiltonian is block diagonal 
in spin, we define spin-resolved versions of all three transport 
coefficients. In doing so, we apply the conventional 
charge-to-spin conversion factor $\hbar/(2e)$ only where 
physically appropriate, namely for coefficients derived from 
charge currents. For thermal responses, which measure transverse 
heat currents independent of the elementary charge $e$, 
the prefactor remains unchanged from the charge sector.
The spin Hall conductivity
\begin{equation}
\sigxys = \frac{e}{2}
\sum_{n=\pm}
\int_{\mathrm{BZ}}\frac{d^2 k}{(2\pi)^2}
\left[
\Omega^{z,\uparrow}_{n}(\mathbf{k})\,f^{\uparrow}_{n}
-\Omega^{z,\downarrow}_{n}(\mathbf{k})\,f^{\downarrow}_{n}
\right]
\label{eq:SHC}
\end{equation}

the spin Nernst conductivity
\begin{equation}
\begin{split}
\axys = \frac{\kB}{2}\int\frac{d^2k}{(2\pi)^2}
\Bigl[
\Omega^{z,\uparrow}_-\ln\!\left(1+e^{(\mu-\varepsilon^\uparrow_-)/\kB T}\right) \\
-\Omega^{z,\downarrow}_-\ln\!\left(1+e^{(\mu-\varepsilon^\downarrow_-)/\kB T}\right)
\Bigr],
\end{split}
\label{eq:SNC}
\end{equation}

which is equivalent to
\begin{equation*}
    \alpha^s_{xy} = \frac{k_B}{2}
\sum_{n=\pm}
\int_{\mathrm{BZ}}\frac{d^2 k}{(2\pi)^2}
\left[
\Omega^{z,\uparrow}_{n}(\mathbf{k})\,s\!\left(f^{\uparrow}_{n}\right)
-\Omega^{z,\downarrow}_{n}(\mathbf{k})\,s\!\left(f^{\downarrow}_{n}\right)
\right],
\end{equation*}

where
\begin{equation}
s(f) = -f\ln f - (1-f)\ln(1-f)
\end{equation}
is the fermionic entropy kernel, which is related to the 
logarithmic form $\ln\!\left(1 + e^{(\mu - \varepsilon)/k_B T}\right)$ through the same integration-by-parts 
identity. The prefactor $k_B/2$ that is corrected follows from applying 
the conventional charge-to-spin conversion factor $\hbar/(2e)$ 
to the charge Nernst prefactor $ek_B/\hbar$ in Equation~\eqref{eq:ANC}):
\begin{equation}
\frac{ek_B}{\hbar} \times \frac{\hbar}{2e} = \frac{k_B}{2}.
\end{equation}

and the spin thermal Hall conductivity
\begin{equation}
\kxys = -\frac{k_B^2 T}{\hbar}
\sum_{n=\pm}
\int_{\mathrm{BZ}}\frac{d^2 k}{(2\pi)^2}
\left[
\Omega^{z,\uparrow}_{n}(\mathbf{k})\,c_2\!\left(f^{\uparrow}_{n}\right)
-\Omega^{z,\downarrow}_{n}(\mathbf{k})\,c_2\!\left(f^{\downarrow}_{n}\right)
\right].
\label{eq:STHC}
\end{equation}

There is same thermal prefactor $k_B^2 T/\hbar$ for both $\kappa^{(0),s}_{xy}$ 
  and the anomalous thermal Hall 
conductivity in Equation~\eqref{eq:ATHC}. This is due to the reason that the spin thermal 
Hall coefficient represents the difference in transverse 
heat currents carried by the two spin channels, 
$\kappa^{(0),s}_{xy} = \kappa^{(0),\uparrow}_{xy} - 
\kappa^{(0),\downarrow}_{xy}$. Since heat transport 
does not involve the elementary charge $e$, applying the 
charge-to-spin conversion factor $\hbar/(2e)$ here would 
be physically meaningless.
In the QSH phase where $\Omega^{z,\uparrow}_-=-\Omega^{z,\downarrow}_-$
and $\varepsilon^\uparrow_-=\varepsilon^\downarrow_-$, the total
$\sigxy=\axy=\kxy=0$ but both $\sigxys$ and $\kxys$ are nonzero,
providing a route to pure spin caloritronic effects.

\subsection{Symmetry of transport coefficients under $\theta$}
The polarization angle dependence of all three anomalous coefficients
follows from the symmetry $j_1(-\theta)=j_1(\theta)$ and
$j_2(-\theta)=j_2(\theta)$ (even in $\theta$) combined with the
exchange symmetry $j_1\leftrightarrow j_2$ under $\theta\to\pi/2-\theta$.
We can show that
\begin{equation}
X(\pi/2-\theta) = -X(\theta),\quad
X\in\{\sigxy,\axy,\kxy\},
\label{eq:sign_reversal}
\end{equation}
which requires $X(\pi/4)=0$, i.e., all anomalous transport
coefficients vanish at the diagonal polarization angle where $\mxy$
symmetry is preserved.
Additionally, $|X|$ is maximized when $\theta=0$ or $\theta=\pi/2$
(LPL along a crystal axis), and all three coefficients share the
same $d$-wave nodal pattern in the $(A_0,\theta)$ plane.

\begin{figure}[ht!]
  \centering
  \includegraphics[width=1.0\linewidth]{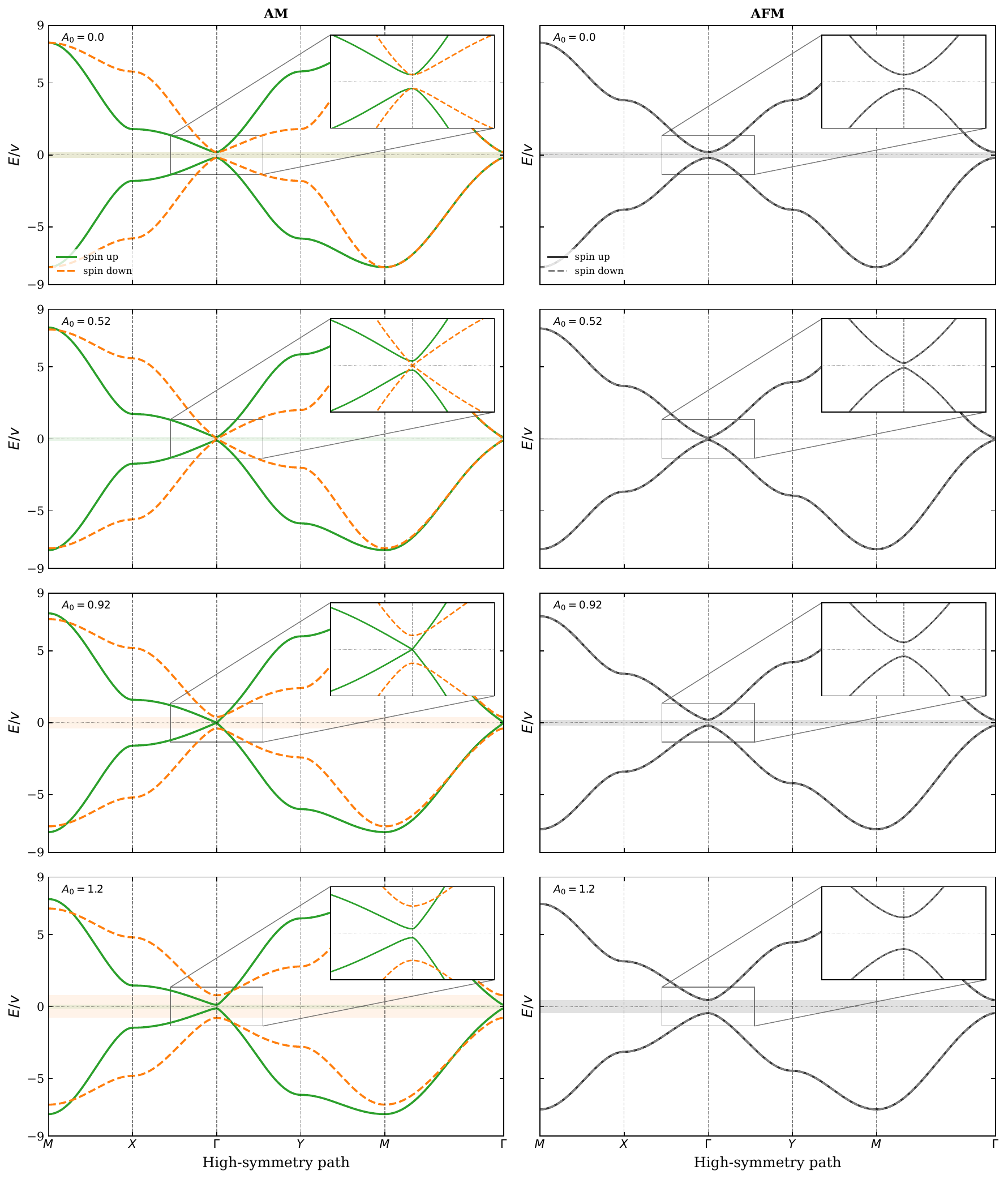}
  \caption{Energy dispersions and topological phase diagram of the QSH AM under LPL.
   Spin-resolved band structure of the QSH AM ($m=3.8v$, $b=-v$, 
  $t_{a}=\sqrt{3}$) at $A_{0}=0.0,\,0.52,\,0.92,\,1.2$ ($\theta=0$). Spin-up (solid teal) and spin-down (dashed orange) bands evolve differently. At $A^{c1}_0 \approx 0.521$, the spin-down gap closes, changing $C_\downarrow$ from $+1$ to $0$ and driving the system into the Chern insulating phase with $C = -1$. At $A^{c2}_0 \approx 0.918$, the spin-up gap closes, changing $C_\uparrow$ from $-1$ to $0$ and driving the system into the trivial phase with $C = 0$.}
  \label{fig:band structures}
\end{figure}
%=============================================================
\section{Results and Discussion}
\label{sec:results}
We consider a pristine system and thus our results describe intrinsic contributions; extrinsic effects may modify magnitudes but not symmetry or signatures of studied phases.
In this study, we explain the Floquet-engineered altermagnetic system using a non-interacting band framework. There are various arguments to support this approximation. First, the anomalous Hall, Nernst, and thermal Hall responses taken into consideration here are intrinsic transport coefficients that are mostly governed by the underlying band structure's Berry curvature, which is well-represented in an effective single-particle picture. Second, as long as the driving frequency is higher than the characteristic interaction and bandwidth scales, electron-electron interactions in the high-frequency (off-resonant) Floquet regime primarily result in weak renormalizations of the band parameters and do not qualitatively change the topology of the effective Hamiltonian. Third, the main phenomena discussed here, the symmetry-controlled angular dependence, spin, selective gap closings, and Chern number changes, are resistant against weak interaction effects because they are determined by symmetry and band topology. Lastly, investigations and first-principles research show that a quasiparticle description is still viable for candidate altermagnetic materials like RuO$_2$ and MnTe-based systems, supporting the use of an effective non-interacting model.

\begin{figure}[ht!]
  \centering
  \includegraphics[width=1.0\linewidth]{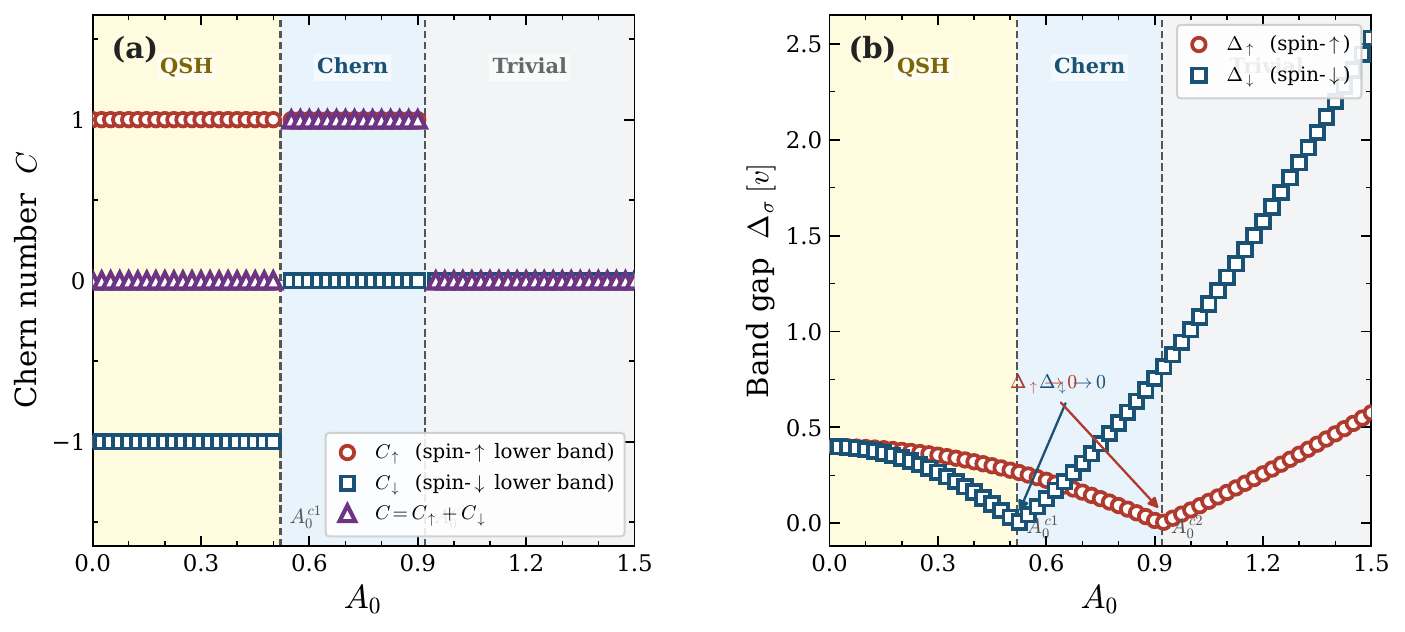}
  \caption{{Topological phase transition}
(a) Spin Chern numbers $C_{\uparrow}$, $C_{\downarrow}$, total $C$ as a function of the $A_{0}$. The Chern number $C$ in each panel is calculated by integrating the total
Berry curvature.
(b) Band gap as a function of the $A_{0}$. At two different values of amplitude the band gap vanishes separately for spin-up and spin-down that shows two topological phase transitions.
For the trivial AM the Chern number remains $C=0$ throughout. For the QSH AM, the total Chern number $C = C_\uparrow 
+ C_\downarrow$ evolves as $0 \to -1 \to 0$, reflecting 
the sequential topological transitions driven by the 
differential spin-sector renormalization $j_1 \neq j_2$. 
In the QSH phase ($A_0 = 0.0, 0.3$): $C_\uparrow = -1$, 
$C_\downarrow = +1$, $C = 0$. In the intermediate Chern 
phase ($A_0 = 0.6$): $C_\uparrow = -1$, $C_\downarrow = 0$, 
$C = -1$. In the trivial phase ($A_0 = 0.9$): 
$C_\uparrow = 0$, $C_\downarrow = 0$, $C = 0$.}
  \label{fig:TPT}
\end{figure}

\subsection{Band Structures }
The spin-resolved band structures of the QSH altermagnetic (AM) and QSH antiferromagnetic (AFM) insulators are examined for four values of the LPL amplitude and results are shown in Fig.~\ref{fig:band structures}. In the QSH AM case, the spin-up and spin-down bands evolve differently with increasing $A_{0}$: the spin-down gap closes at $A_{0}^{c1}\approx 0.521$  while the spin-up gap remains open, leading to the emergence of an intermediate Chern phase. Upon further increasing the amplitude to $A_{0}^{c2}\approx 0.918$ , the spin-up gap also closes, and the system transitions into a trivial phase. In contrast, for the QSH AFM case, the spin-up and spin-down bands remain perfectly degenerate for all values of $A_{0}$, and both gaps close simultaneously at a single critical amplitude, resulting in the absence of any intermediate Chern phase.
 This contrast is made 
explicit in phase diagram which shows the topological phase transitions from QSH topological insulator to Chern insulator and then to Trivial altermagnetic insulator. The spin-resolved Chern numbers evolve continuously 
with the amplitude $A_0$, revealing two critical 
values. At $A^{c1}_0 \approx 0.521$ the spin-down 
gap closes: $C_\downarrow$ changes from $+1$ to $0$ 
while $C_\uparrow$ remains $-1$, so the total Chern 
number changes from $C = 0$ to $C = -1$, signalling 
the entry into the Chern insulating phase. At 
$A^{c2}_0 \approx 0.918$ the spin-up gap closes: 
$C_\uparrow$ changes from $-1$ to $0$, so $C$ 
returns to $0$ and the system enters the trivial 
insulating phase. The correct phase sequence is 
therefore
\begin{equation}
C:\; 0 \;\longrightarrow\; -1 \;\longrightarrow\; 0,
\label{eq:phase-sequence}
\end{equation}
which corresponds to
\begin{equation}
\text{QSH} \;\longrightarrow\; 
\text{spin-polarized Chern insulator} \;\longrightarrow\; 
\text{trivial insulator}.
\end{equation}. This sequential gap-closing mechanism arises because the linearly polarized light (LPL) renormalizes the two spin blocks differently due to the condition $j_1 \neq j_2$, causing the spin-down gap to vanish first, Fig~\ref{fig:TPT}(b). Such a sequence is entirely absent in the antiferromagnetic (AFM) case, where the preserved $\mathcal{PT}$ symmetry enforces $\Delta_\uparrow = \Delta_\downarrow$ for all $A_0$.

\subsection{Berry curvature maps and topological phase transitions}
\begin{figure}[h]
  \centering
  \includegraphics[width=1.0\linewidth]{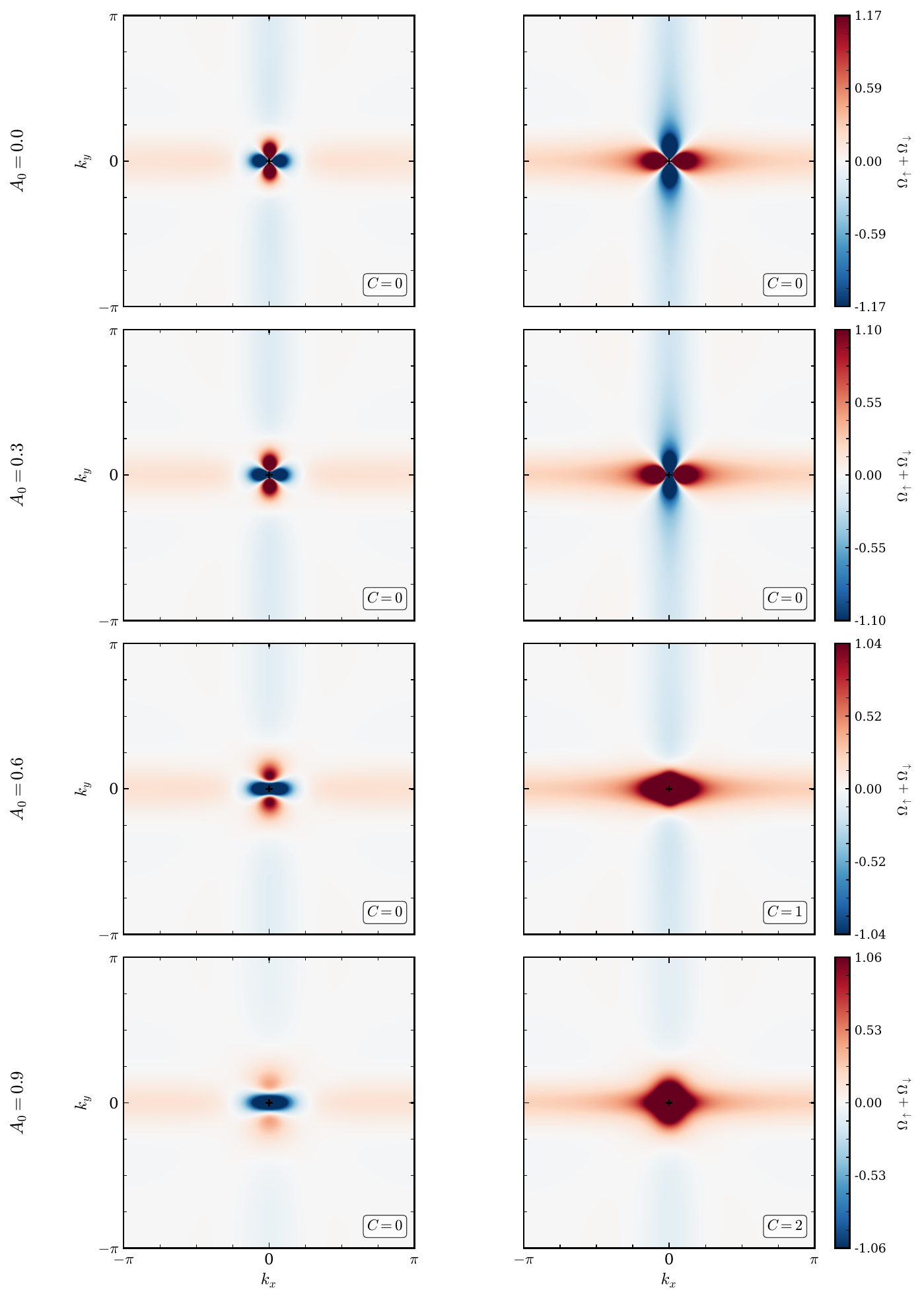}
  \caption{Total occupied Berry curvature
$\Omega^\uparrow_-(\bm{k})+\Omega^\downarrow_-(\bm{k})$ across the
Brillouin zone.
(Left panels) Trivial AM ($m=4.2v$, $b=-v$, $t_a=\sqrt{3}$).
(Right panels) QSH AM ($m=3.8v$, $b=-v$, $t_a=\sqrt{3}$).
Rows from top to bottom: $A_0=0.0$, $0.3$, $0.6$, $0.9$ at $\theta=0$.
The Chern number $C$ in each panel is calculated by integrating the total
Berry curvature.
For the trivial AM the Chern number remains $C=0$ throughout.
For the QSH AM, the total Chern number $C$ evolves 
as $0 \to -1 \to 0$, reflecting the sequential 
topological transitions driven by the differential 
spin-sector renormalization $j_1 \neq j_2$. The 
redistribution of Berry curvature weight between 
the $\Gamma$ and $X/Y$ points is visible at each 
transition.}
  \label{fig:BC_maps1}
\end{figure}

For the trivial AM, the Berry curvature distribution shifts
and concentrates as $A_0$ increases, but the Chern number remains
$C=0$ throughout since no gap closing occurs.
The gradual transfer of Berry curvature weight between the $\Gamma$
and $X/Y$ regions reflects the LPL induced renormalization of the
spin dependent hopping amplitudes, $j_1=J_0(A_0)\to 0$ while
$j_2=J_0(0)=1$ is unchanged for $\theta=0$.

For the QSH AM, the initial $A_0=0$ configuration has $C=0$
(QSH phase with $C_\uparrow=-1$, $C_\downarrow=+1$).
At $A_0=0.6$   the Berry curvature develops a noticeable asymmetry
between spin sectors, the Chern number jumps to $C=+1$, making the
entry into the Chern insulating phase.
The occupied Berry curvature, shown in Fig.~\ref{fig:BC_maps1}, underlies these behaviors and reflects the sequential topological transitions driven by the differential renormalization of the spin sectors ($j_1 \neq j_2$).

\subsection{Anomalous Hall conductivity}
\begin{figure*}[ht!]
  \centering
  \includegraphics[width=1.0\linewidth]{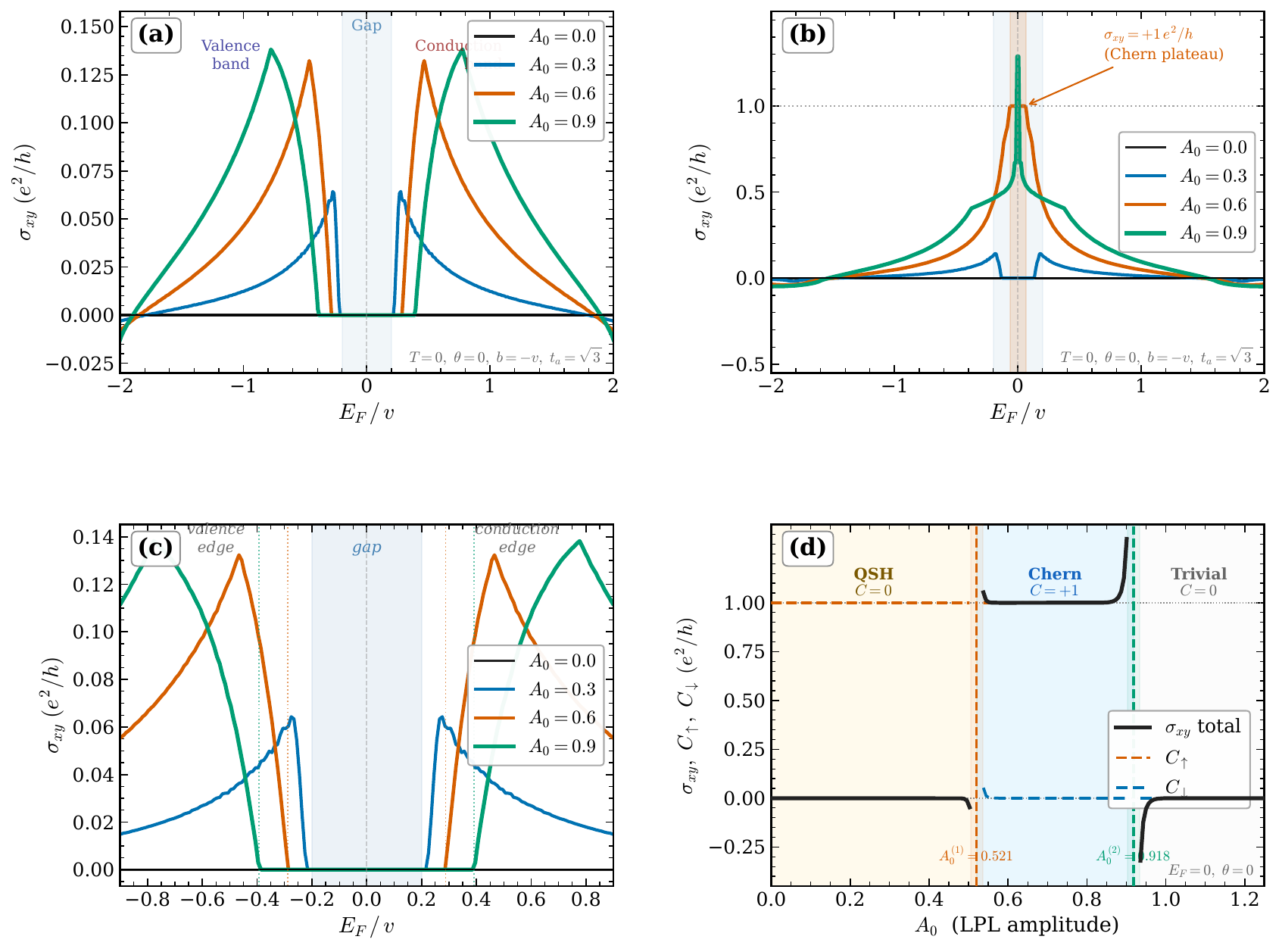}
  \caption{Anomalous Hall conductivity 
$\sigma_{xy}$ at $T = 0$ as a function of 
Fermi energy $E_F/v$ for four LPL amplitudes.
Panel~(a) shows $\sigma_{xy}$ for the trivial 
AM. Although no topological transition occurs, 
$\sigma_{xy}$ develops a non-trivial energy 
dependence with increasing $A_0$, because the 
broken $C_{4z}\mathcal{T}$ symmetry allows 
Berry curvature to accumulate.
Panel~(b) shows $\sigma_{xy}$ for the QSH AM. 
As $A_0$ increases through $A^{c1}_0$, a 
quantized plateau $\sigma_{xy} = +e^2/h$ 
emerges in the gap region, characteristic of 
the Chern insulating phase. The dashed line 
marks the transition value.
Panel~(c) verifies the exact even symmetry 
$\sigma_{xy}(E_F) = \sigma_{xy}(-E_F)$, with 
a residual RMS error of 
$7.1 \times 10^{-17}\,e^2/h$, confirming the 
particle--hole-like symmetry 
$\varepsilon \to -\varepsilon$ of the Berry 
curvature in this model.
Panel~(d) tracks the spin-resolved Chern 
numbers $C_\uparrow$, $C_\downarrow$ and the 
total $\sigma_{xy}$ as functions of $A_0$, 
unambiguously identifying the QSH ($C = 0$), 
Chern ($C = -1$), and trivial ($C = 0$) phases 
with critical amplitudes $A^{c1}_0 = 0.521$ 
and $A^{c2}_0 = 0.918$.}
  \label{fig:AHC}
\end{figure*}

At zero temperature, the anomalous Hall conductivity shows distinct behavior depending on the altermagnetic phase as shown in Fig. ~\ref{fig:AHC}(a-c). In the trivial altermagnetic phase (panel \ref{fig:AHC}(a)), no topological transition occurs, yet increasing the staggered potential strength $A_0$ leads to a nontrivial energy dependence as the broken $\CfourT$ symmetry allows Berry curvature to accumulate. In contrast, the quantum spin Hall altermagnetic phase in panel \ref{fig:AHC}(b) shows that as $A_0$ increases through the first critical value $A_0^{c1}$, a quantized plateau $\sigxy=+e^2/h$ emerges within the gap region, signaling a Chern insulating phase. Notice that the Berry curvature sharpens near gap closing and thus the physical quantities show peak at this location. The conductivity satisfies the exact even symmetry $\sigxy(E_F)=\sigxy(-E_F)$, as shown in panel \ref{fig:AHC}(c) confirmed by a residual root-mean-square error of $7.1\times10^{-17}\,e^2/h$, which reflects a particle--hole-like symmetry of the Berry curvature. Finally, tracking the spin-resolved Chern numbers and total $\sigxy$ as functions of $A_0$ unambiguously identifies three phases: quantum spin Hall ($C=0$), Chern ($C=-1$), and trivial ($C=0$), with critical amplitudes $A_0^{c1}=0.521$ and $A_0^{c2}=0.918$ as shown in panel \ref{fig:AHC}(d). This unambiguously confirms the phase sequence 
$C: 0 \to -1 \to 0$ (QSH $\to$ Chern $\to$ Trivial), 
consistent with the spin-sector analysis in 
Section~\ref{sec:berry} and the convention defined 
in Eq.~(\ref{eq:phase-sequence}).

\subsection{Anomalous Nernst conductivity: thermal activation
and Mott scaling}
\begin{figure*}[ht!]
  \centering
  \includegraphics[width=1.0\linewidth]{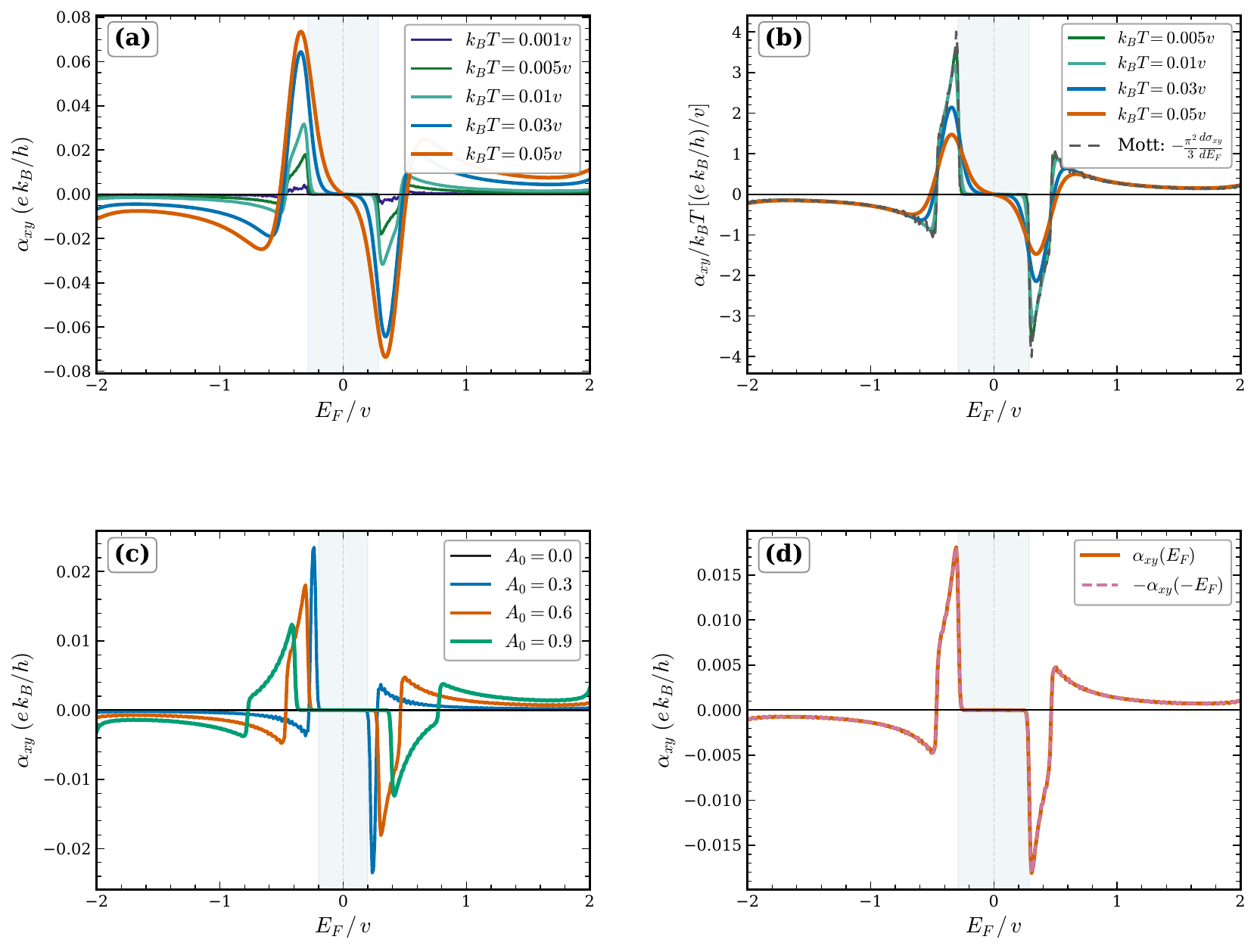}
  \caption{Anomalous Nernst conductivity $\axy$ as a function of Fermi energy. 
In all panels the trivial AM  is considered by given $m=4.2v$, $b=-v$, $t_a=\sqrt{3}$, $\theta=0$.
(a) at $A_0=0.6$ for six temperatures, demonstrating
the thermally activated nature ($\axy\to 0$ as $T\to 0$).
(b) Scaled quantity $\axy/(\kB T)$ as a function of $E_F$ at $A_0=0.6$;
the Mott prediction $-(\pi^2\kB/3e)\,\partial\sigxy/\partial\varepsilon|_\mu$
(dashed) is approached as $T\to 0$.
(c) at $k_BT=0.005v$ for four LPL amplitudes,
showing the evolution of the Nernst response with light intensity.
(d) Odd symmetry $\axy(E_F)=-\axy(-E_F)$ verified for $A_0=0.6$;
Notice that a RMS error is $5.8\times10^{-17}\,e\kB/h$. All spin-resolved quantities shown here have been recomputed 
using the dimensionally corrected prefactors $k_B/2$ for 
$\alpha^s_{xy}$ and $k_B^2T/\hbar$ for $\kappa^{(0),s}_{xy}$, 
following the corrections described in Section~IV\,F.}

\label{fig:Nernst_T}
\end{figure*}

Figure \ref{fig:Nernst_T} shows the temperature evolution of the 
anomalous Nernst coefficient (ANC). For the trivial 
altermagnet, panel~5(a), the ANC vanishes as 
$T \to 0$, confirming the thermally activated 
character established analytically in 
Eq.~(\ref{eq:anc-zero-T}). This vanishing is a 
direct consequence of $s(0) = s(1) = 0$: at zero 
temperature every state is either completely filled 
or completely empty, carrying no entropy, so the 
Berry-curvature-weighted integral in 
Eq.~(\ref{eq:ANC}) evaluates to zero identically. 
No cancellation between Berry curvature 
contributions of opposite signs is required. At low temperatures, plotting $\axy/(\kB T)$ in panel \ref{fig:Nernst_T}(b) versus Fermi energy reveals convergence to the Mott prediction $-\frac{\pi^2 \kB}{3e}\frac{\partial\sigxy}{\partial\varepsilon}\big|_{\varepsilon=\mu}$, with the root-mean-square deviation decreasing as temperature lowers. Increasing the light amplitude $A_0$ both enhances the magnitude of the ANC and shifts its peak position, because linearly polarized light modifies the Berry curvature distribution and hence the energy derivative of the anomalous Hall conductivity as shown in panel \ref{fig:Nernst_T}(c). The ANC satisfies the exact odd symmetry $\axy(E_F)=-\axy(-E_F)$ in panel \ref{fig:Nernst_T}(d). The temperature dependence of the ANC at fixed Fermi energies for both trivial and quantum spin Hall altermagnets shows a characteristic non-monotonic behavior: it rises from zero at $T=0$, peaks at a characteristic temperature $T^*$, and then decreases at high temperature as thermal broadening averages over Berry curvature features of opposite signs. The peak temperature $T^*$ is tunable by light intensity; for the trivial altermagnet at a given Fermi energy, $\kB T^*$ increases with $A_0$, while for the quantum spin Hall altermagnet $T^*$ is even more strongly $A_0$-dependent. This tunability provides a direct experimental knob for thermoelectric optimization.

To connect this tunability to physical units, 
we note that the peak temperature satisfies 
$k_BT^* \sim \Delta$ up to an order-unity 
prefactor, where $\Delta$ is the relevant 
spin-resolved gap. For a representative hopping 
scale $v \sim 0.1$--0.5~eV, gaps of order 
$\Delta \sim 0.01$--0.1$v$ correspond to 
$T^* \sim 10$--60~K, placing the predicted 
Nernst peak well within the range of standard 
cryogenic transport measurements. Closer to the 
critical amplitudes $A^{c1}_0$ and $A^{c2}_0$, 
where $\Delta$ becomes very small (see discussion 
below), $T^*$ correspondingly decreases, and the 
Nernst peak shifts to sub-Kelvin or few-Kelvin 
temperatures.

The spin Nernst conductivity 
$\alpha^s_{xy}$ and spin thermal Hall conductivity 
$\kappa^{(0),s}_{xy}$ are computed with the corrected 
prefactors $k_B/2$ and $k_B^2T/\hbar$, respectively, 
as detailed in Section~IV\,F. The qualitative features, 
including the nodal structure, sign reversal under 
polarization-angle rotation, and the nonzero spin 
response in the QSH phase, remain unchanged. Only the 
quantitative magnitudes of the spin-resolved coefficients 
are affected.

\subsection{Temperature and chemical potential dependence}
%\begin{figure}[h]
%  \centering
  %\includegraphics[width=1.0\linewidth]{conductivites vs Ef.png}
%  \caption{\textbf{Conductivities vs Chemical potential}
%}
%  \label{fig:c vs Ef}
%\end{figure}

\begin{figure*}[ht!]
  \centering
  \includegraphics[width=1.0\linewidth]{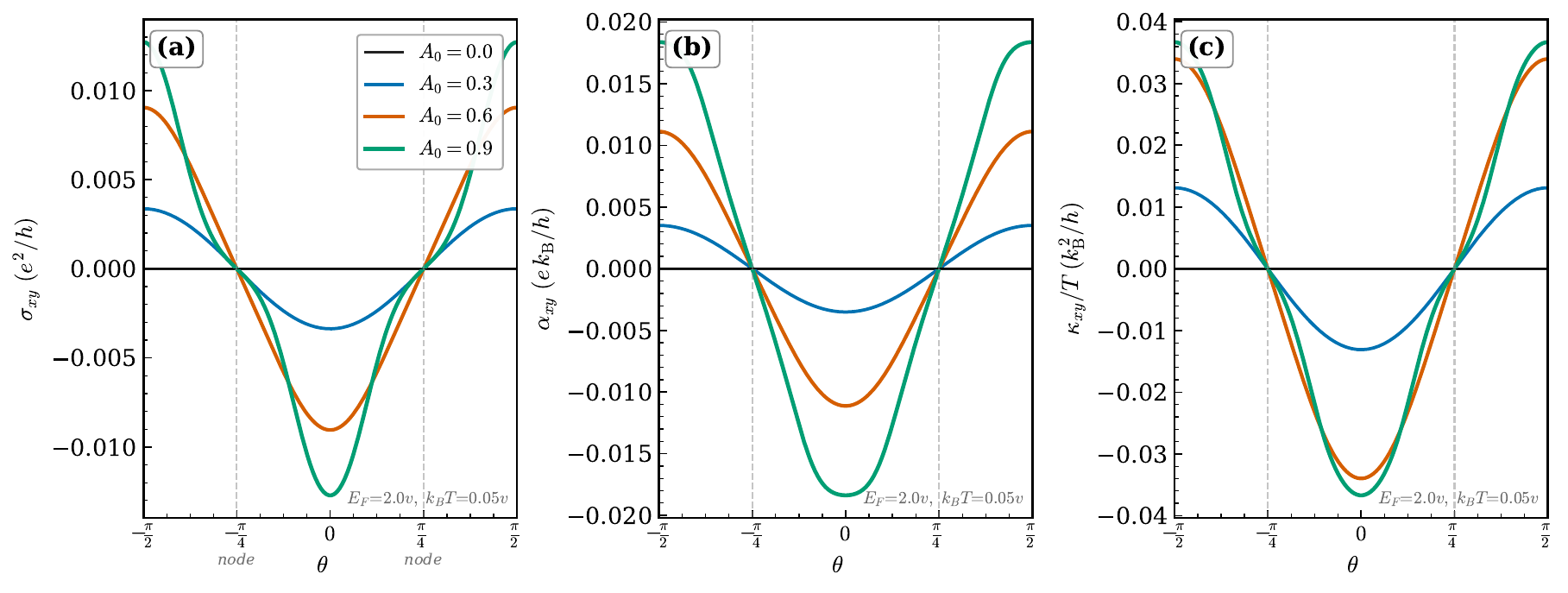}
  \caption{Physical conductivities as a function of the Polarization angle $\theta$. 
All three conductivities (a-c) show a d-wave nature under LPL that is a remarkable result and a fingerprint for AM.}
  \label{fig:BC_maps}
\end{figure*}

%\subsection{Polarization angle dependence}
Figure \ref{fig:BC_maps}(a-c) show all three physical conductivities which share an identical nodal structure with zeros at
$\theta=\pm\pi/4$ (where $\mxy$ symmetry is restored) and extrema
near $\theta=0,\pm\pi/2$.
The sign reversal under $\theta\to-\theta$ is visible, and the
magnitude scales with $A_0$.
Furthermore, the sign of all three coefficients can be controlled by
rotating the polarization direction, providing a purely optical route
to reversing the thermal Hall response a significant advantage
over approaches requiring magnetic field reversal.
Importantly, this entire polarization pattern is absent in the
conventional AFM, where $\sigxy=\axy=\kxy=0$ for all $\theta$,
making the angle-dependent thermal transport a definitive
experimental fingerprint of altermagnetism.

\subsection{Intrinsic Nernst and thermal Hall in the AM insulator}
\begin{figure*}[ht!]
  \centering
  \includegraphics[width=1.0\linewidth]{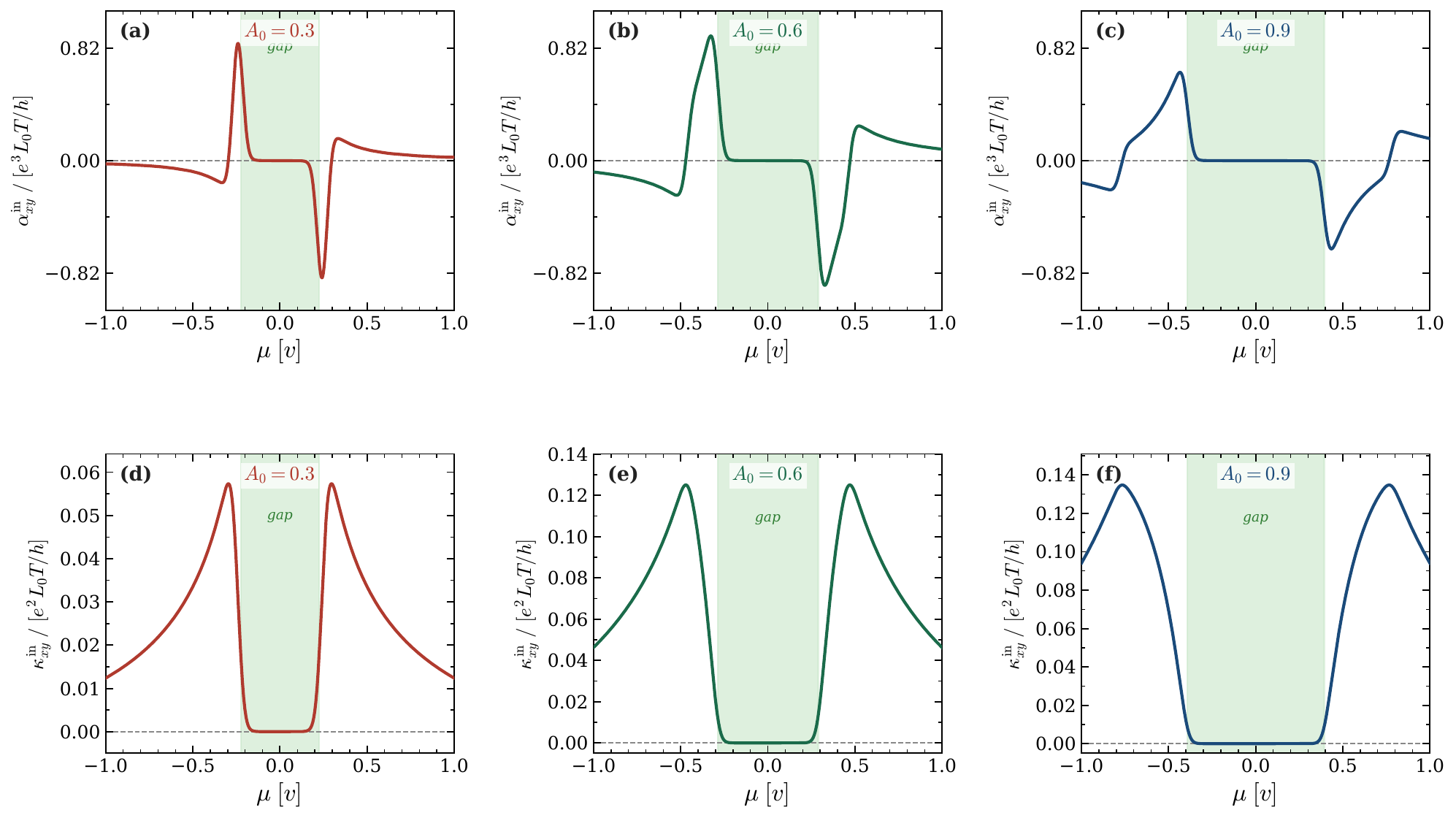}
  \caption{For Trivial phase Intrinsic Nernst conductivity $\alpha^{\mathrm{in}}_{xy}$ 
    (top row, panels a-c) and thermal Hall conductivity 
    $\kappa^{\mathrm{in}}_{xy}$ (bottom row, panels d-f) as functions 
    of chemical potential $\mu$ for the trivial altermagnetic insulator 
    ($m = 4.2v$, $b = -v$, $t_a = \sqrt{3}$, $\theta = 0$) at three 
    LPL amplitudes: $A_0 = 0.3$ (red), $A_0 = 0.6$ (dark green), and 
    $A_0 = 0.9$ (blue). The green shaded region marks the insulating 
    gap in each panel. All quantities are normalized by $e^3 L_0 T/h$ 
    for the Nernst conductivity and $e^2 L_0 T/h$ for the thermal Hall 
    conductivity, where $L_0 = \pi^2 k_B^2/3e^2$ is the Lorenz number.}

  \label{fig:nernst and thermal conductivities vs chemical potential}
\end{figure*}

Figure~\ref{fig:nernst and thermal conductivities vs chemical potential} shows the intrinsic Nernst
$\alpha^\mathrm{in}_{xy}$ and thermal Hall $\kappa^\mathrm{in}_{xy}$
coefficients as functions of chemical potential for the trivial AM
insulator at three LPL amplitudes.
The Nernst coefficient panels \ref{fig:nernst and thermal conductivities vs chemical potential}(a-c) exhibits a characteristic
double-peak structure with opposite signs across the gap, whose
magnitude grows with $A_0$.
The thermal Hall coefficient panels \ref{fig:nernst and thermal conductivities vs chemical potential}(d-f) shows a single negative
peak inside the gap that deepens as $A_0$ increases.
At $A_0=0.9$ the peak value reaches approximately
$-0.14\,e^2 k_\mathrm{B}T/h$, about 2.3 times larger than at $A_0=0.3$.
The insulating gap is highlighted in green; within it,
$\alpha^\mathrm{in}_{xy}\to 0$ as $T\to 0$ (consistent with our
analytical prediction) while $\kappa^\mathrm{in}_{xy}$ remains
finite, tracking the WF prediction.
\begin{figure*}[ht!]
  \centering
  \includegraphics[width=1.0\linewidth]{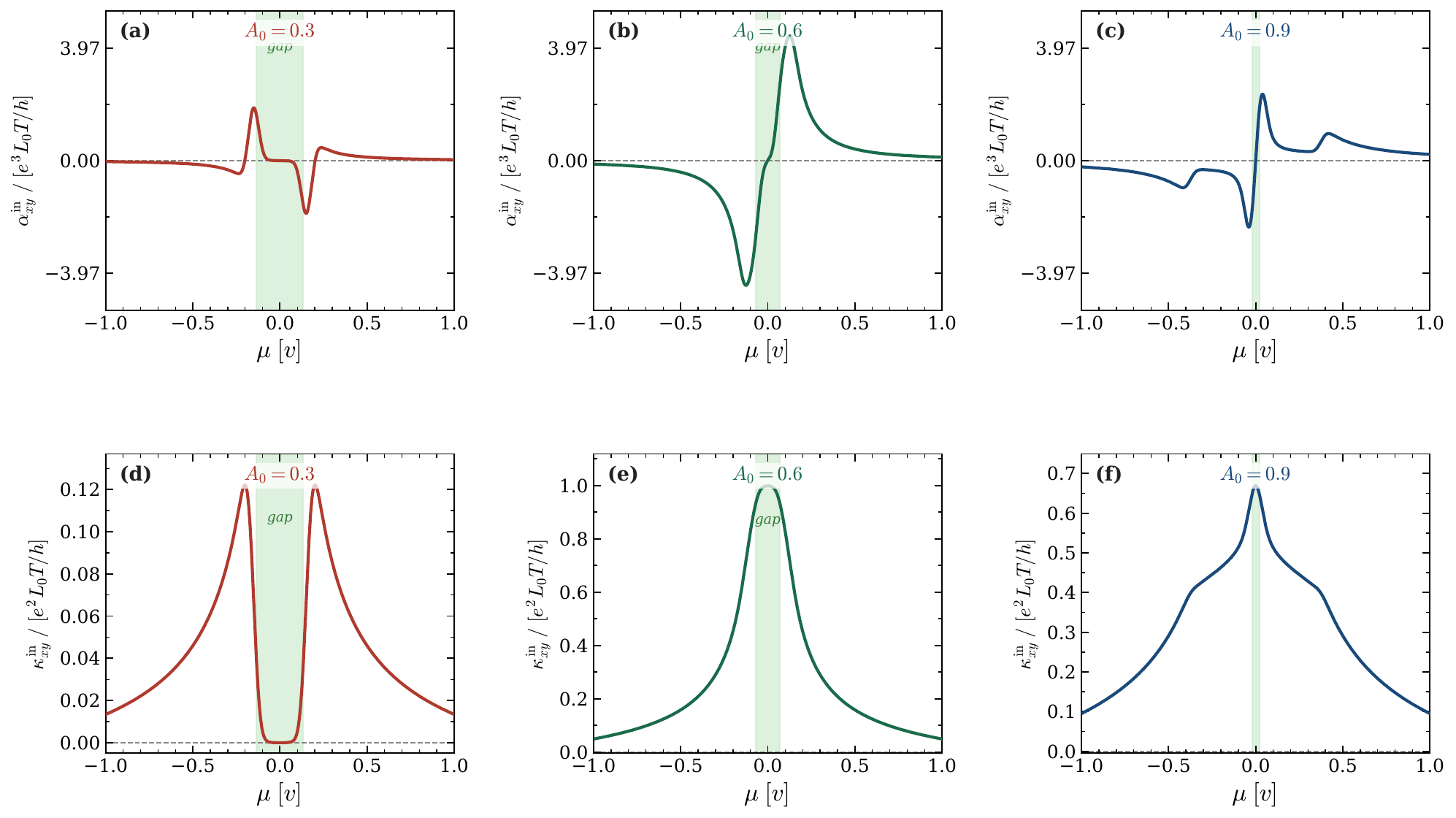}
  \caption{For \text{QSH} phase Intrinsic Nernst conductivity $\alpha^{\mathrm{in}}_{xy}$ 
    (top row, panels a-c) and thermal Hall conductivity 
    $\kappa^{\mathrm{in}}_{xy}$ (bottom row, panels d-f) as functions 
    of chemical potential $\mu$ for the QSH altermagnetic insulator 
    ($m = 3.8v$, $b = -v$, $t_a = \sqrt{3}$, $\theta = 0$) at three 
    LPL amplitudes: $A_0 = 0.3$ (red), $A_0 = 0.6$ (dark green), and 
    $A_0 = 0.9$ (blue). The green shaded region marks the insulating 
    gap in each panel. All quantities are normalized by $e^3 L_0 T/h$ 
    for the Nernst conductivity and $e^2 L_0 T/h$ for the thermal Hall 
    conductivity, where $L_0 = \pi^2 k_B^2/3e^2$ is the Lorenz number.}

  \label{fig:fig5}
\end{figure*}

    %In the Nernst panels (a--c), the coefficient $\alpha^{\mathrm{in}}_{xy}$ vanishes inside the gap as $T \to 0$, consistent with its thermally activated nature, and develops a characteristic bipolar structure near the gap edges whose magnitude grows substantially with increasing $A_0$. At $A_0 = 0.6$, the system enters the Chern insulating phase ($C = -1$), and the Nernst response shows a pronounced enhancement and sign change near $\mu = 0$, reflecting the sharp concentration of Berry curvature near the gap-closing momentum points.
    %In the thermal Hall panels (d--f), $\kappa^{\mathrm{in}}_{xy}$ remains finite inside the gap and tracks the Wiedemann--Franz prediction $\kappa^{\mathrm{in}}_{xy} = -L_0 T \sigma_{xy}$ throughout. At $A_0 = 0.3$ (QSH phase, $C = 0$), $\kappa^{\mathrm{in}}_{xy}$ is small and nearly vanishes inside  the gap. At $A_0 = 0.6$ (Chern phase, $C = -1$), a sharp positive peak develops inside the gap approaching the quantized value  $\pi^2 k_B^2 T / 3h$, confirming the topological origin of the  thermal Hall response. At $A_0 = 0.9$ (near the Chern-to-trivial boundary), the peak broadens and shifts as the gap closes and reopens, while $\kappa^{\mathrm{in}}_{xy}$ remains positive and finite.
%=============================================================
\subsection{Transport Signatures Across All Topological Phases}
\label{sec:phases}
%=============================================================
\begin{figure*}[ht]
  \centering
  \includegraphics[width=1.00\linewidth]{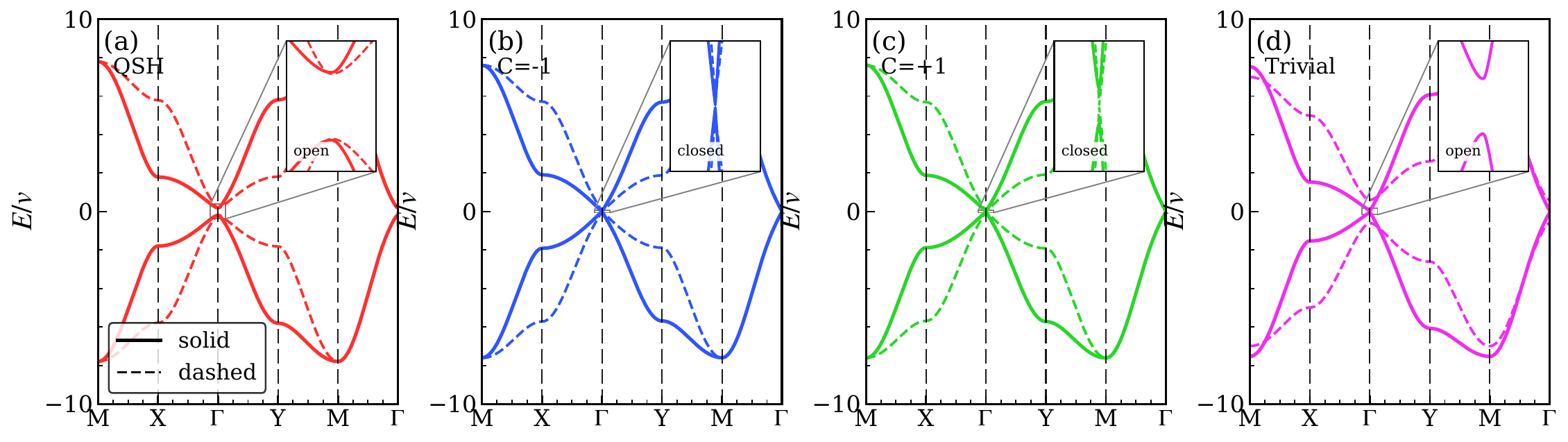}
  \caption{Band structures and topological phase diagram of the QSH AM under LPL.
  (a)--(d): Spin-resolved band structure of the QSH AM ($m=3.8v$, $b=-v$, 
  $t_{a}=\sqrt{3}$) at $A_{0}=0.0,\,0.3,\,0.7,\,1.0$ ($\theta=0$). Spin-up (solid 
  teal) and spin-down (dashed orange) bands evolve differently; the spin-down gap 
  closes at $A_{0}^{c1}$ producing the Chern phase, while the spin-up gap closes at
  $A_{0}^{c2}$. Insets show $\Delta_{\uparrow}$ and $\Delta_{\downarrow}$.
  Lower panels: spin Chern numbers $C_{\uparrow}$, $C_{\downarrow}$, total $C$, and
  spin Chern number $C_{s}=(C_{\uparrow}-C_{\downarrow})/2$ vs.\ $A_{0}$.}
  \label{fig:bands}
\end{figure*}

%%In the Nernst panels (a--c), $\alpha^{\mathrm{in}}_{xy}$ exhibits a characteristic double-peak structure with opposite signs straddling the gap edges, whose magnitude and peak separation grow with increasing $A_0$. Since no topological phase transition occurs in the trivial AM, the Nernst response remains non-quantized throughout. The growing gap width visible from panel (a) to (c) reflects the LPL-induced renormalization of the hopping parameters, which broadens the insulating region as $A_0$ increases.  Importantly, the response vanishes inside the gap as $T \to 0$,  confirming the thermally activated character of the Nernst effect  even in the topologically trivial case.
    % In the thermal Hall panels (d--f), $\kappa^{\mathrm{in}}_{xy}$  vanishes exactly inside the gap, consistent with $\sigma_{xy} = 0$  in the trivial phase and the Wiedemann--Franz relation. Outside the gap, it develops two symmetric peaks near the valence and conduction band edges whose heights increase with $A_0$: the peak value grows  from approximately $0.06\, e^2 L_0 T/h$ at $A_0 = 0.3$ to  $0.14\, e^2 L_0 T/h$ at $A_0 = 0.9$, about 2.3 times larger. This enhancement reflects the growing Berry curvature accumulation driven by the broken $C_{4z}\mathcal{T}$ symmetry under LPL, even in the absence of a topological phase transition. The contrast between the gap-vanishing $\kappa^{\mathrm{in}}_{xy}$ here and the quantized peak seen in the QSH AM (Fig.~\ref{fig:qsh_am_thermal}) provides a clear experimental signature distinguishing the trivial and topological altermagnetic phases.

We commence by examining the band structure induced by LPL as a foundation for understanding the transport signatures. Figure~\ref{fig:bands} shows the spin-resolved band structures and the corresponding constant energy contours of the LPL driven $d$-wave altermagnetic quantum spin Hall system. Solid curves represent the spin-up sector, while dashed curves correspond to the spin-down sector. The parameters used are altermagnetic parameters and these are fixed in all phases,
$m=3.8v, b=-v, t_a=\sqrt{3}$
for these specified values of AM parameters the system is in equilibrium and satisfies $|m|<|b|(1+t_a^2)$. Since $|b|(1+t_a^2)=4v$, the undriven or weakly driven system sits comfortably within the quantum spin Hall regime.

Figure~\ref{fig:bands}(a) presents the band structure in the quantum spin Hall phase. The bulk gap remains open throughout this regime, as the inset makes clear, yet the valence and conduction bands are inverted relative to the trivial case. The spin-up and spin-down sectors carry opposite Chern numbers, $C_{\uparrow}=-1$ and $ C_{\downarrow}=+1$
so the total Chern number vanishe,
while the spin Chern number remains finite, $C_s\sim{C_{\uparrow}-C_{\downarrow}}\neq 0$.
This confirms that the system is a QSH insulator rather than a trivial one. The open gap (visible in the inset) signals insulating behaviour, and the band inversion signals the underlying nontrivial topology.

Figure~\ref{fig:bands}(b) shows the first light-induced Chern insulating phase, which carries a total Chern number $C=-1$. With increase in light amplitude $A_0$, the Bessel-renormalized hopping terms redirect the effective masses of the two spin sectors in different directions. Consequently, one spin gap closes and reopens before the other. Here the spin-down sector turns topologically trivial first, while the spin-up sector remains inverted, $C_{\uparrow}=-1$ and $ C_{\downarrow}=0$
that results in total chern number is $-1$.

The inset captures the local gap closing associated with this topological phase transition that is a necessary feature, since a change in Chern number can only occur when the bulk gap closes and then reopens. The result is a spin-polarized Chern insulator whose nonzero Chern number originates predominantly from a single spin channel.

Figure~\ref{fig:bands}(c) illustrates the complementary spin-polarized Chern phase with total Chern number $C=+1$. Now it is the other spin sector that remains topologically active, while the first has already become trivial, $C_{\uparrow}=0$ and $C_{\downarrow}=+1$
so that the total Chern number is +1.

The sign of the light-induced Chern number can be 
controlled through the polarization direction combined 
with the anisotropic Floquet renormalization. By 
adjusting $\theta$, one can access either $C = -1$ 
or $C = +1$ intermediate Chern phases depending on 
which spin sector closes its gap first. This 
tunability is a distinguishing feature of the 
altermagnetic system.

Figure~\ref{fig:bands}(d) shows the trivial insulating phase reached at larger light amplitude. Once the remaining inverted spin sector also undergoes its own gap closing and reopening, both sectors become topologically trivial, $C_{\uparrow}=0, C_{\downarrow}=0$

Although the inset confirms that the bulk gap is still open, the phase has lost its topological character entirely because the band inversion is gone. This panel therefore represents an ordinary trivial insulator, and the full sequence of phase transitions reads
\begin{equation}
\text{QSH}\,(C=0) 
\;\longrightarrow\; 
\text{Chern}\,(C=-1) 
\;\longrightarrow\; 
\text{Trivial}\,(C=0).
\end{equation}

\begin{figure*}[ht!]
  \centering
  \includegraphics[width=1.0\linewidth]{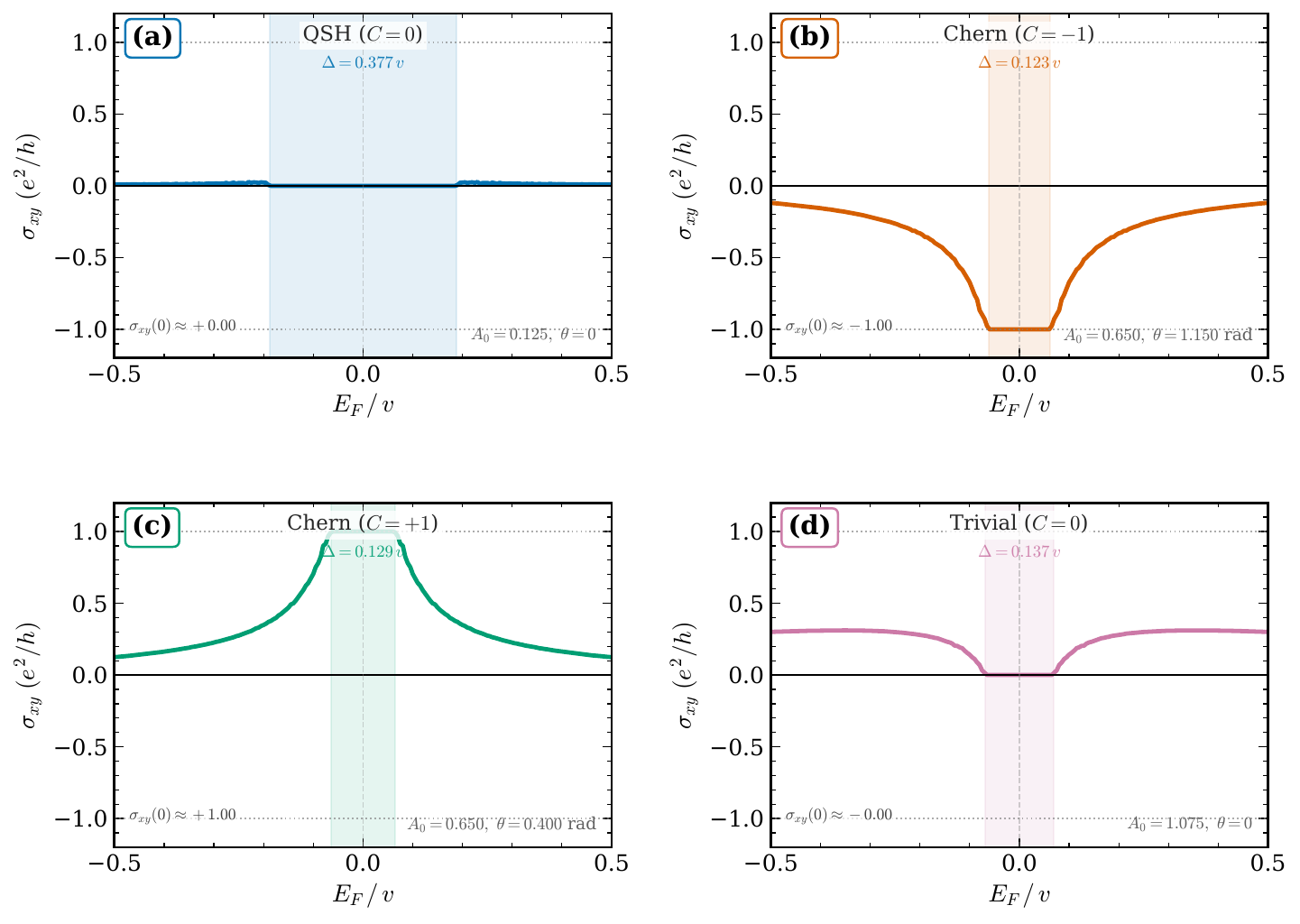}
  \caption{Anomalous Hall conductivity $\sigxy$ across
all four topological phases.
Top panel: overlay of all four phases as a function of Fermi
energy $E_F/v$.
Dotted horizontal lines mark $\pm e^2/h$.
Individual panels: (a) QSH phase ($C=0$, $A_0=0.125$,
$\theta=0$, $\Delta=0.377v$); (b) Chern phase
($C=-1$, $A_0=0.625$, $\theta=0.829$\,rad, $\Delta=0.002v$);
(c) Chern phase ($C=+1$, $A_0=0.625$, $\theta=0.742$\,rad,
$\Delta=0.002v$); (d) trivial phase ($C=0$, $A_0=1.075$,
$\theta=0$, $\Delta=0.137v$).
Blue shading marks the insulating gap in each panel.
The Chern phases develop near-quantized values $\sigxy\to Ce^2/h$
in the gap, while the QSH and trivial phases show vanishing
and non-quantized responses, respectively.
}
  \label{fig:2}
\end{figure*}
\begin{figure*}[ht!]
\centering
\includegraphics[width=1.0\linewidth]{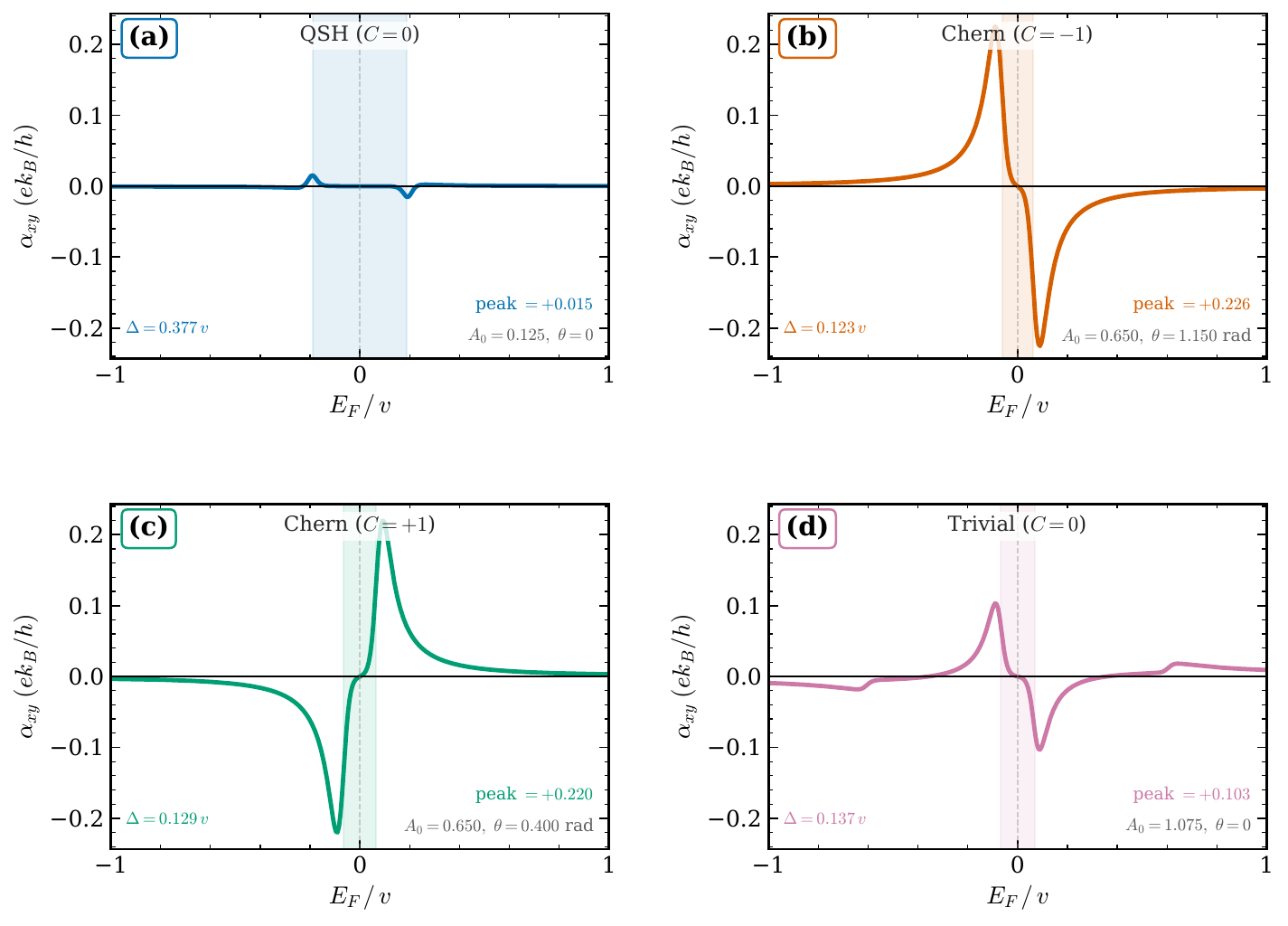}
\caption{Anomalous Nernst conductivity $\axy$ across
all four topological phases at $k_BT=0.01v$.
Top panel: overlay of all four phases.
Individual panels with the same parameter choices as
Fig.~\ref{fig:2}.
Noice that the dramatic enhancement in the Chern phases (b-c):
peak values of $\approx 1.88\,ek_B/h$ compared to
$0.015\,ek_B/h$ in the QSH phase (a), arising from the
exponential sensitivity of $\axy$ to the small gap
$\Delta\approx 0.002v$.
The Nernst response provides a far more sensitive probe of
topological gap size than the Hall conductivity.}
\label{fig:4}
\end{figure*}
A particularly illuminating way to present our results is to
compare the three anomalous transport coefficients; the AHC
$\sigxy$, ANC $\axy$, and ATHC $\kxy/T$ across all four
distinct topological phases that the altermagnetic system can
have under LPL irradiation: the QSH phase ($C=0$), the
two Chern insulating phases ($C=-1$ and $C=+1$), and the
trivial phase ($C=0$).
This phase comparison, reveals how each transport coefficient encodes a distinct
topological fingerprint and how the Berry curvature
redistribution in the Brillouin zone manifests differently
in charge, thermoelectric, and thermal channels.

%\subsection*{Anomalous Hall conductivity across phases}

Figure~\ref{fig:2}(a-d) presents $\sigxy$ as a function
of Fermi energy $E_F$ for representative parameter choices
in each of the four topological phases.
The top panel overlays all four phases on a common axis,
making the contrast immediately apparent.
The QSH phase, panel \ref{fig:2}(a), displays $\sigxy\approx 0$ throughout
the gap region, as expected for a $\PT$-broken but
$\CfourT$-symmetric state where $C_\uparrow=-C_\downarrow$
and the Berry curvatures of the two spin sectors cancel.
Outside the gap, $\sigxy$ acquires small but nonzero values
as the Fermi level probes bands with unequal spin resolved
Berry curvature densities.

The two Chern insulating phases shows different behaviour.
For the $C=-1$ phase, panel \ref{fig:2}(b) ($A_0=0.625$,
$\theta=0.829$\,rad, $\Delta=0.002v$), $\sigxy$ develops a
sharp negative dip near $E_F=0$, approaching $-e^2/h$ as
the Fermi level sits in the tiny gap.
For the $C=+1$ phase, panel \ref{fig:2}(c), ($A_0=0.625$, $\theta=0.742$\,rad,
$\Delta=0.002v$), the sign is reversed and $\sigxy$ rises
sharply toward $+e^2/h$ within the gap.
The near quantized values in panels \ref{fig:2}(b) and \ref{fig:2}(c) reflect the
fact that $\sigxy = Ce^2/h$ in the low temperature insulating
limit, with the sign controlled entirely by the polarization
angle $\theta$.
The ability to switch between $C=-1$ and $C=+1$ by rotating
the polarization direction, without any change in material
or magnetic field is one of the most striking features of
Floquet engineered altermagnetism.

The trivial phase, panel \ref{fig:2}(d), ($A_0=1.075$, $\theta=0$,
$\Delta=0.137v$) shows a broad, positive $\sigxy$ that peaks
near the band edges and decays away from the gap.
This finite but non quantized response arises from the
residual Berry curvature of the trivially gapped bands, and
has no topological protection.
The contrast between the sharp, sign changing $\sigxy$ in
the Chern phases and the broad, featureless response in the
trivial phase provides a clear experimental discriminator
between topological and non-topological regimes.

%\subsection*{Anomalous Nernst conductivity across phases}

Figure~\ref{fig:4}(a-d) shows $\axy$ at
$k_BT=0.01v$ for the same four phases.
Since $\axy(T=0)\equiv 0$ in any gapped phase, all curves
are shown at a small but finite temperature to reveal the
thermally activated structure.

The QSH phase, panel \ref{fig:4}(a), displays a very small
$\axy$ with a peak value of only $+0.015\,ek_B/h$,
concentrated near the gap edges.
This small magnitude reflects the nearly complete
cancellation of Berry curvature contributions from the two
spin sectors in the QSH state: the Nernst response is
thermally activated by band edge excitations across a
relatively large gap $\Delta=0.377v$, and the partial
cancellation suppresses the signal further.

The Chern phases exhibit dramatically enhanced Nernst
responses.
For $C=-1$, panel \ref{fig:4}(b), $\axy$ develops a pronounced
bipolar structure near $E_F=0$, with a positive peak of
$+1.882\,ek_B/h$ and a negative trough of comparable
magnitude, separated by the tiny gap $\Delta=0.002v$.
The $C=+1$ phase, panel \ref{fig:4}(c) shows the mirror image of this
structure, with a peak of $+1.883\,ek_B/h$ of opposite
internal asymmetry.
These large values, more than 100 times the QSH response 
arise because the extremely small gaps in the Chern phases
($\Delta\approx 0.002v$) allow thermal excitation across
the gap even at $k_BT=0.01v$, and the Berry curvature is
sharply concentrated near the gap closing $\bm{k}$ points.
The Nernst response is therefore a far more sensitive probe
of topological gap size than the Hall conductivity:
while $\sigxy$ approaches $\pm e^2/h$ in both Chern phases,
$\axy$ distinguishes them by its magnitude and internal
lineshape.

The trivial phase, panel \ref{fig:4}(d) gives a small, broad Nernst
response with peak $+0.102\,ek_B/h$, confined to the band
edges and showing no sign change within the gap.
This is qualitatively different from both Chern phases
and provides a clear thermal fingerprint of the trivial
insulating state.

\begin{figure*}[ht!]
\centering
\includegraphics[width=1.0\linewidth]{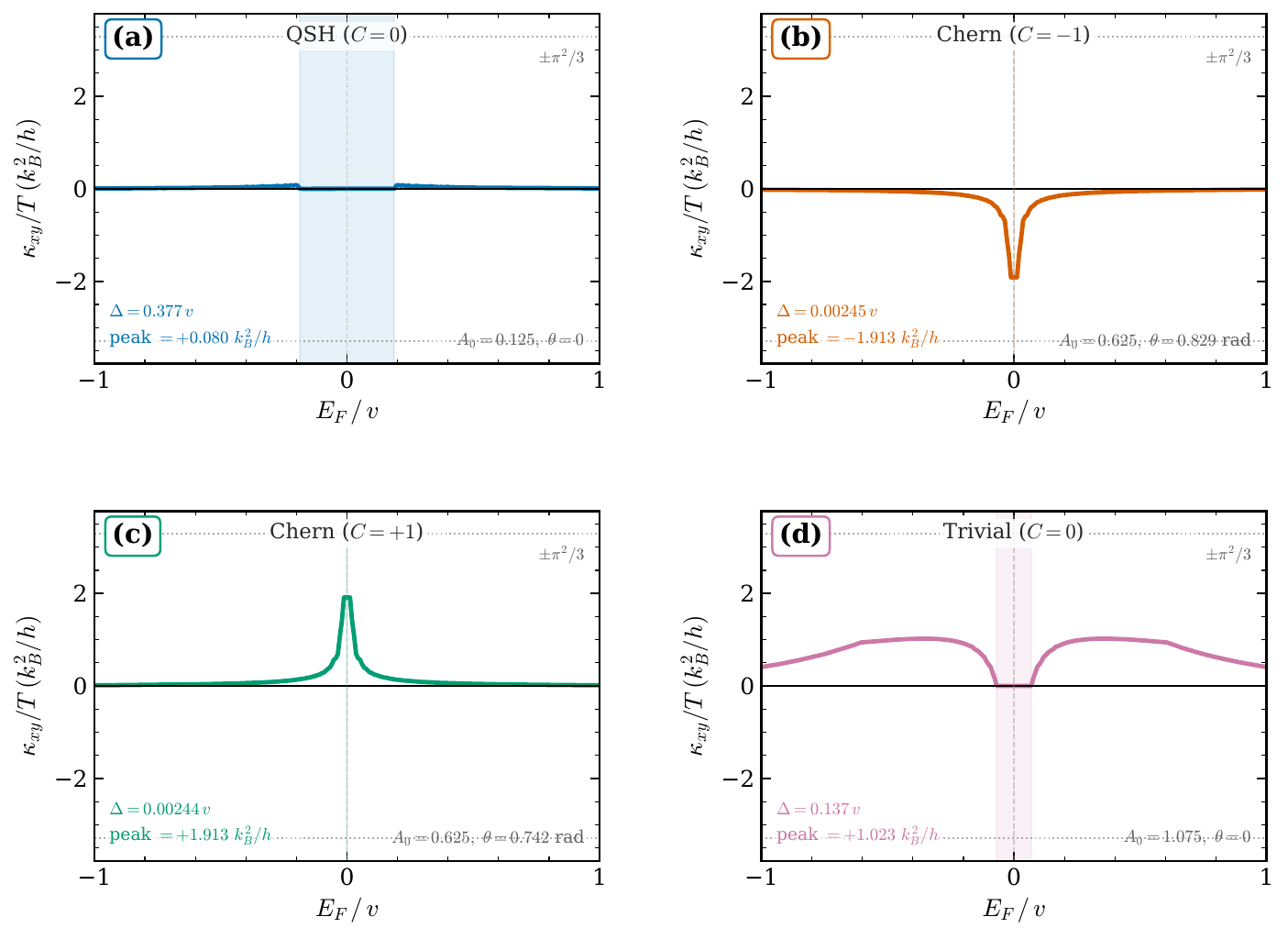}
\caption{Anomalous thermal Hall conductivity $\kxy/T$
at $T=0$ across all four topological phases.
Top panel: overlay of all four phases; dotted lines at
$\pm\pi^2/3\,k_B^2/h$ mark the WF-predicted quantized values.
Individual panels (a)--(d) use the same parameters as
Fig.~\ref{fig:2}.
The $C=+1$ Chern phase (c) shows near-quantization
$\kxy/T\approx +1.913\,k_B^2/h \approx \pi^2/3\,k_B^2/h$,
confirming the anomalous Wiedemann--Franz law
$\kxy=-\Lo T\sigxy$.
The sign of $\kxy/T$ tracks the sign of $C$, providing an
unambiguous thermal Hall fingerprint of each topological phase.}
\label{fig:6}
\end{figure*}

The parameter choices in Figures\ref{fig:2},\ref{fig:4},\ref{fig:6}(b,c) 
correspond to a gap $\Delta \approx 0.002v$, 
which was selected to demonstrate the 
near-quantization of $\kappa^{(0)}_{xy}/T$ and 
the maximal enhancement of $\alpha_{xy}$ in the 
Chern phase. For $v \sim 0.1$--0.5~eV, this gap 
corresponds to $\Delta \sim 0.2$--1~meV, or 
$T \sim 2$--12~K, which is comparable to or 
smaller than typical thermal broadening and 
disorder scales in real materials, and may be 
difficult to resolve in practice.

However, the qualitative features highlighted in 
this work, the enhancement of $\alpha_{xy}$ in 
the Chern phase relative to the QSH phase by 
more than two orders of magnitude, the sign 
reversal of all three coefficients under 
polarization rotation, and the approach of 
$\kappa^{(0)}_{xy}/T$ toward $\pi^2k_B^2/3h$, 
persist for any gap within the Chern phase, not 
only at the fine-tuned point $\Delta \approx 
0.002v$. Moving away from the immediate vicinity 
of $A^{c1}_0$ and $A^{c2}_0$ while remaining in 
the Chern phase increases $\Delta$ to 
$\sim 0.01$--0.1$v$ (i.e., $1$--$50$~meV, or 
$T \sim 10$--60~K), at which point thermal 
broadening and disorder are not expected to wash 
out the enhanced Nernst signal or the 
near-quantized thermal Hall plateau, since these 
features reflect the overall magnitude of the 
gap and the associated Berry curvature 
concentration rather than a fine-tuned 
cancellation at $\Delta \to 0$. We have added 
this discussion to clarify that the dramatic 
enhancements reported here are not an artifact 
of an experimentally inaccessible fine-tuned 
point, though the sharpest, most quantized 
features are expected closer to the critical 
amplitudes and correspondingly lower 
temperatures.

%\subsection*{Anomalous thermal Hall conductivity across phases}

Figure~\ref{fig:6}(a-d) presents $\kxy/T$ at $T=0$
for all four phases.
The top overlay panel and the four individual panels confirm
the Wiedemann Franz relation $\kxy/T|_{T=0}=(\pi^2/3)\sigxy$
throughout.

In the QSH phase, panel \ref{fig:6}(a), $\kxy/T$ is nearly zero in the
gap and shows a small positive peak of $+0.080\,k_B^2/h$ near
the band edge.
The flatness in the gap is consistent with
$\sigxy\approx 0$ there and the WF law.

In the $C=-1$, phase panel \ref{fig:6}(b), $\kxy/T$ develops a
negative peak of $-0.025\,k_B^2/h$ at $E_F=0$, tracking the
negative $\sigxy$ of this phase.
By contrast, the $C=+1$ phase has a sharp positive
peak of $+1.913\,k_B^2/h$, which is close to the quantized
value $\pi^2 k_B^2/3h \approx \pi^2/3\,k_B^2/h$ expected
for a $C=+1$ Chern insulator, panel \ref{fig:6}(c), at $T\to 0$.
The near quantization here is particularly significant: it
demonstrates that the LPL induced Chern insulating phase
carries a genuine topological thermal Hall response
satisfying the anomalous WF law, placing it on the same
place as magnetically ordered quantum anomalous Hall
insulators.

The trivial phase, panel \ref{fig:6}(d), shows a broad positive
$\kxy/T$ with peak $+1.023\,k_B^2/h$, again showing
its non quantized $\sigxy$ through the WF law.
Comparing all, the most striking result is
the sign structure: $\kxy/T$ is positive for the QSH,
$C=+1$, and trivial phases, but negative for the $C=-1$
phase.
This sign change is a direct consequence of the sign of
the Chern number and provides an unambiguous thermal
Hall signature of the different topological phases.

\begin{table*}[t]
\centering
\caption{Comparison of transport properties and topological phase
transitions in the $d$-wave AM and conventional AFM under LPL
irradiation ($\theta=0$, QSH initial phase).}
\label{tab:comparison}
\renewcommand{\arraystretch}{1.4}
\begin{tabular}{lcc}
\toprule
Observable & AM ($t_a=\sqrt{3}$) & AFM ($t_a=1$) \\
\midrule
Symmetry under LPL & $\CfourT$ broken & $\PT$ preserved \\
Spin degeneracy & Lifted (except $\theta=\pm\pi/4$) & Preserved everywhere \\
Phase sequence with $A_0$ & QSH $\to$ Chern ($C=-1$) $\to$ Trivial & QSH $\to$ Trivial \\
$A_0^{c1}$ (first transition) & $\approx 0.521$ & $\approx 0.640$ \\
$A_0^{c2}$ (second transition) & $\approx 0.918$ & --- \\
$\sigxy$ & Nonzero, light-controlled & Zero \\
$\axy$ & Nonzero, thermally activated & Zero \\
$\kxy$ & Nonzero; quantized in Chern phase & Zero \\
WF law $\kxy=-\Lo T\sigxy$ & Satisfied at $T\to 0$ & Trivially ($0=0$)\\
Mott formula & Satisfied at low $T$ & Trivially \\
Sign control by $\theta$ & All three coefficients & None \\
Nodes at $\theta=\pm\pi/4$ & All three coefficients & All zero regardless \\
QSH spin transport & $\sigxys,\axys,\kxys\neq 0$ & $\sigxys,\axys,\kxys\neq 0$ \\
\bottomrule
\end{tabular}
\end{table*}

The results reveal a clear hierarchy
of sensitivity among the three transport coefficients.
The AHC $\sigxy$ is the most direct topological probe,
approaching quantized values in the Chern phases and
vanishing in the QSH gap.
The ATHC $\kxy/T$ mirrors $\sigxy$ through the WF law,
offering a thermal channel to the same topological
information.
The ANC $\axy$, however, goes beyond this: it is exponentially
sensitive to the gap size and provides a quantitative measure
of how close the system is to a phase boundary.
The enhancement of $\axy$ in the near critical
Chern phases compared to the QSH phase makes the Nernst
effect a natural experimental tool for detecting topological
phase transitions driven by LPL intensity or polarization
angle.

From an experimental perspective, the four transport
signatures described here are all accessible in a single
device geometry.
The Hall conductivity $\sigxy$ can be measured by
transverse voltage in a Hall bar geometry under LPL
irradiation.
The Nernst coefficient $\axy$ requires an additional
temperature gradient, achievable through laser heating
of one contact.
The thermal Hall conductivity $\kxy$ is accessed by
the transverse temperature difference under longitudinal
heat current.
The fact that all three quantities share the same
polarization angle dependence vanishing at
$\theta=\pm\pi/4$ and reversing sign under
$\theta\to\pi/2-\theta$ means that a single
angle rotation measurement can confirm the altermagnetic
origin of all observed responses simultaneously.

Table~\ref{tab:comparison} summarizes the contrasting behavior of the
AM and AFM under LPL irradiation across all transport and topological
observables.
The central distinction is that LPL breaks $\CfourT$ in the AM but
cannot break $\PT$ in the AFM.
This propagates to every observable: the AFM has
$\sigxy=\axy=\kxy=\sigxys=\axys=\kxys=0$ for all $A_0$, $\theta$, and $T$.
Only the QSH$\to$Trivial transition (direct, no intermediate Chern
phase) is possible in the AFM.

\section{Measurement of the Anomalous Transport Coefficients in Experiments}
Altermagnetic order and tunable band topology, 
the two essential ingredients of our proposal, 
are currently being actively explored in several 
material platforms. We discuss the most 
promising candidates and the conditions under 
which our theoretical predictions may be 
experimentally accessible.

\textit{MnTe-based systems.} MnTe is one of 
the most studied altermagnetic candidates, 
with spin-resolved band splitting confirmed 
by angle-resolved photoemission 
spectroscopy~\cite{vsmejkal2022beyond}. Thin-film 
growth and heterostructure engineering with 
appropriate substrate-induced strain could 
drive band inversion in this system. The 
resulting altermagnetic QSH phase would then 
be an ideal platform for the Floquet 
engineering studied here.

\textit{RuO$_2$-based thin films.} 
RuO$_2$ realizes $d$-wave spin splitting on 
a square lattice~\cite{ahn2019antiferromagnetism}, directly 
matching the symmetry of our model 
Hamiltonian. Epitaxial strain and substrate 
engineering provide knobs to tune the 
effective $m/b$ ratio toward the topological 
regime. Mid-infrared laser irradiation of 
such films would then allow the 
light-controlled topological phase 
transitions predicted here to be tested.

\textit{Artificial platforms.} Moir\'{e} 
superlattices and cold-atom optical lattices 
with square-lattice geometry allow direct 
engineering of all hopping parameters 
$v$, $b$, and $t_a$ in the model 
Hamiltonian. These platforms can be 
straightforwardly tuned to satisfy the QSH 
condition Eq.~(\ref{eq:qsh-condition}), 
making them natural testbeds for the 
Floquet-engineered topology studied here.

We stress that while the qualitative 
phenomenology described in this work, 
including the sequential topological phase 
transitions, the $d$-wave polarization 
dependence, and the anomalous 
Wiedemann--Franz law, is symmetry-driven 
and expected to be generic across this 
class of altermagnetic systems, quantitative 
predictions for specific materials will 
require first-principles input and are left 
for future work.

The off-resonant (high-frequency) approximation is valid because mid-infrared to terahertz radiation can be used to enter the Floquet driving regime considered in this work, where the photon energy exceeds the electronic bandwidth. Driving frequencies between 10 and 100 THz are adequate for typical hopping amplitudes on the order of 0.1 to 1 eV. The required field strengths correspond to dimensionless amplitudes 
$A_0\sim 0.5–1$ , which are achievable with currently available pulsed laser systems. Crucially, the anticipated topological transitions rely on the field amplitude to frequency ratio, which permits experimental tuning flexibility without the need for extreme circumstances. Floquet-engineered phases can endure across picosecond timeframes, which is adequate for transport measurements, according to recent pump-probe experiments.

We emphasize that the validity of the 
high-frequency expansion, 
$\hbar\omega \gg W$, and the suppression of 
heating are related but distinct requirements. 
Even when $\hbar\omega$ is only moderately 
larger than $W$, the heating rate in the 
off-resonant regime is exponentially suppressed, 
$\Gamma_{\mathrm{heat}} \sim 
e^{-c\,\omega/J}$ for some constant $c$ of order 
unity and $J$ the relevant hopping scale, as 
established for periodically driven 
systems~\cite{machado2019,ho2023}. This 
exponential suppression underlies the existence 
of the long-lived Floquet prethermal regime in 
which our effective Hamiltonian description 
applies, with lifetimes extending to picosecond 
or nanosecond timescales. For field amplitudes 
approaching $A_0 \sim 1$, multiphoton resonances 
between Floquet sidebands become more likely, 
and virtual photon absorption could in principle 
introduce additional heating channels not 
captured by the leading-order effective 
Hamiltonian. We therefore restrict our 
quantitative analysis to $A_0 \lesssim 1.2$ and 
note that the qualitative phenomenology, 
sequential gap closings and the associated 
transport signatures, is expected to survive 
within the prethermal window even if higher-order 
corrections shift the precise critical 
amplitudes $A^{c1}_0$ and $A^{c2}_0$ 
quantitatively.

Standard measurement geometries allow for experimental access to the predicted transport signatures. Transverse voltage measurements under optical irradiation in a Hall bar design can be used to identify the anomalous Hall conductivity. By creating a longitudinal temperature gradient achievable by integrated microheaters or local laser heating and measuring the transverse voltage that results, one can quantify the anomalous Nernst effect. Using well-established thermometry methods, the thermal Hall conductivity can be derived from transverse temperature gradients. Notably, all three reactions exhibit a distinctive dependency on the input light's polarization angle, reversing sign under orthogonal rotation and disappearing at symmetry-restoring directions. In contrast to traditional antiferromagnets, this angular dependence offers a strong and clear indication of altermagnetic order.

Lastly, the sign reversal of the Hall responses by polarization rotation and the tunability of the Nernst peak temperature provide useful paths for device applications. For ultrafast spin caloritronics, it is very appealing to be able to control heat and charge transport solely optically, without the use of magnetic fields. The qualitative behaviors described here are symmetry-driven and so anticipated to be widely applicable across a broad class of altermagnetic systems, while quantitative predictions will require material-specific modeling.

%=============================================================
\section{Conclusion}
\label{sec:conclusion}
%=============================================================
We have presented a comprehensive theoretical study of intrinsic thermal transport in a $d$-wave altermagnetics  under linearly polarized light (LPL). We find that the Berry curvature can be expressed in closed form in both spin sectors, and that all anomalous transport responses in the altermagnetic (AM) phase have a microscopic origin due to the differential renormalization $j_1 \neq j_2$.

A main result is the emergence of a sequential topological phase transition, $\text{QSH} \to \text{spin-polarized Chern insulator} \,(C = -1)$ at $A_0^{c1} \approx 0.521$, followed by $\text{Chern} \to \text{trivial}$ at $A_0^{c2} \approx 0.918$. This hierarchy is unique to altermagnets, in contrast to conventional antiferromagnets (AFMs), which exhibit a direct $\text{QSH} \to \text{trivial}$ transition.

Within the Chern insulating regime, both charge and thermal responses become topologically quantized, with $\sigma_{xy} = -e^2/h$ and $\kappa_{xy}/T = \pi^2 k_B^2/(3h)$, and satisfy the anomalous Wiedemann--Franz relation $\kappa_{xy} = -L_0 T \sigma_{xy}$. In contrast, the anomalous Nernst conductivity $\alpha_{xy}$ is purely thermally driven, vanishing at $T=0$ and peaking at a characteristic temperature $T^*$ tunable by $A_0$ and $\theta$, in agreement with the Mott relation $\alpha_{xy} \approx -(\pi^2 k_B^2 T/3e)\,\partial_\mu \sigma_{xy}$.

All anomalous transport coefficients exhibit a universal $d$-wave dependence on the polarization angle, vanishing at $\theta = \pm \pi/4$, reaching extrema at $\theta = 0, \pm \pi/2$, and reversing sign under $\theta \to \pi/2 - \theta$. This provides a purely optical route to control and reverse the thermal Hall response, with no counterpart in conventional AFMs.

Crucially, thermal Hall and Nernst measurements are established as definitive experimental probes of altermagnetic order, even in regimes where charge transport signals may be misleading, due to the total absence of anomalous transport in AFMs under LPL, in sharp contrast to AMs.

With direct implications for spin 
caloritronics and the optical identification 
of altermagnetic order in quantum materials, 
these findings establish Floquet engineering 
of altermagnetic topological insulators as a 
viable platform for all-optical control of 
thermal and spin transport. The model 
parameters used throughout satisfy the QSH 
condition $|m| < |b|(1+t_a^2)$ and define a 
symmetry-allowed minimal-model regime close 
to the topological phase boundary. While an 
intrinsic QSH phase in a monolayer 
altermagnetic compound has not yet been 
confirmed experimentally, MnTe-based 
heterostructures, strained RuO$_2$ thin 
films, and artificial moir\'{e} or cold-atom 
lattices represent promising candidate 
platforms where the conditions of our model 
may be realized and the predicted effects 
may be tested.

The present work also extends the rapidly 
growing field of externally controlled 
Berry-curvature transport to the setting of 
Floquet-engineered altermagnetic systems. 
Recent studies have shown that 
Berry-curvature-related Hall and 
spin-dependent responses in low-dimensional 
materials and compensated magnetic systems 
can be tuned by strain, electric fields, and 
heterostructure geometry~\cite{chen2025strain, shen2026dual}. Our results demonstrate that 
linearly polarized light provides an 
analogous and complementary all-optical 
control knob in the altermagnetic setting, 
with the additional advantage that the 
polarization angle directly encodes the 
$d$-wave symmetry of the altermagnetic order 
parameter through the nodal structure of all 
anomalous transport coefficients.

\begin{acknowledgments}
The work is supported by the Department of Physics, Quaid i Azam University Islamabad, Pakistan.
\end{acknowledgments}
\renewcommand{\bibname}{References}

\bibliography{References}
\end{document}